\documentclass[preprintnumbers,prd,twocolumn,showpacs,floatfix,preprintnumbers,letterpaper,amsmath,amssymb,nofootinbib,superscriptaddress]{revtex4}
\usepackage{graphicx}
\usepackage{epsfig}
\usepackage{bm}
\usepackage{amsfonts}
\usepackage{color}

\newcommand{\la}{\lambda}

\newcommand{\bear}{\begin{eqnarray}}

\newcommand{\eear}{\end{eqnarray}}

\newbox\pippobox

\def\6{\partial}

\def\a{\alpha}

\def\sq
\def\a{\alpha}

\def\dn{\Delta N_{\nu}}

\newcommand {\lla} {\ {\raise-.5ex\hbox{$\buildrel<\over\sim$}}\ }

\def\be{\begin{equation}}
\def\ee{\end{equation}}
\def\ba{\begin{eqnarray}}
\def\ea{\end{eqnarray}}
\def\w{\omega}
\renewcommand{\(}{\left(}
\renewcommand{\)}{\right)}

\renewcommand{\]}{\right]}

\begin{document}

\title{Ho\v{r}ava-Lifshitz cosmology with generalized Chaplygin gas}

\author{Amna Ali} \email{amnaalig@gmail.com}
\affiliation{Center For Theoretical Physics, Jamia Millia Islamia,
New Delhi 110025, India}

\author{Sourish Dutta}
\email{sourish.d@gmail.com} \affiliation{Department of Physics and School of Earth and Space Exploration,
Arizona State University, Tempe, AZ 85287}

\author{Emmanuel N. Saridakis}
 \email{msaridak@phys.uoa.gr}
 \affiliation{College of Mathematics and Physics,\\ Chongqing University of Posts and
Telecommunications, Chongqing, 400065, P.R. China }

\author{Anjan A. Sen} \email{anjan.ctp@jmi.ac.in}
\affiliation{Center For Theoretical Physics, Jamia Millia Islamia,
New Delhi 110025, India}

\begin{abstract}
We investigate cosmological scenarios of generalized Chaplygin gas
in a universe governed by Ho\v{r}ava-Lifshitz gravity. We consider
both the detailed and non-detailed balance versions of the
gravitational background, and we include the baryonic matter and
radiation sectors. We use observational data from  Type Ia
Supernovae (SNIa), Baryon Acoustic Oscillations (BAO), and Cosmic
Microwave Background (CMB), along with requirements of Big Bang
Nucleosynthesis (BBN), to constrain the parameters of the model,
and we provide the corresponding likelihood contours. We deduce
that the present scenario is compatible with observations.
Additionally, examining the evolution of the total
equation-of-state parameter, we find in a unified way the
succession of the radiation, matter, and dark energy epochs,
consistently with the thermal history of the universe.
\end{abstract}


\maketitle

\section{Introduction}

Recently  Ho\v{r}ava proposed a power-counting
renormalizable theory with consistent ultra-violet (UV) behavior
\cite{hor2,hor1,hor3,hor4}. Although presenting an infrared (IR)
fixed point, namely General Relativity, in the  UV the theory
exhibits an anisotropic, Lifshitz scaling between time and space.
Due to these novel features, there has been a large amount of
effort in examining and extending the properties of the theory
itself
\cite{Cai:2009ar,Cai:2009dx,Nishioka:2009iq,Charmousis:2009tc,Li:2009bg,Visser:2009fg,
Sotiriou:2009bx,Germani:2009yt,Chen:2009bu,Bogdanos:2009uj,Kluson:2009rk,Afshordi:2009tt,Myung:2009ur,Alexandre:2009sy,
Blas:2009qj,Capasso:2009fh,Chen:2009vu,Kluson:2009xx,Kiritsis:2009vz,Garattini:2009ns,Kluson:2010aw,Son:2010qh,Carloni:2010nx,Eune:2010qk,Wang:2010mw,Gullu:2010wb}.
Additionally, application of Ho\v{r}ava-Lifshitz gravity as a
cosmological framework gives rise to Ho\v{r}ava-Lifshitz
cosmology, which proves to lead to interesting behavior
\cite{Calcagni:2009ar,Kiritsis:2009sh}. In particular, one can
examine specific solution subclasses
\cite{Lu:2009em,Minamitsuji:2009ii,Wu:2009ah,Cho:2009fc,Boehmer:2009yz,Momeni:2009au,Cai:2010ud,Huang:2010rq},
the phase-space behavior
\cite{Carloni:2009jc,Leon:2009rc,Myung:2009if,Bakas:2009ku,Myung:2010qg},
the gravitational wave production
\cite{Mukohyama:2009zs,Park:2009gf,Park:2009hg,Myung:2009ug}, the
perturbation spectrum
\cite{Mukohyama:2009gg,Piao:2009ax,Chen:2009jr,Gao:2009ht,Cai:2009hc,Wang:2009yz,Kobayashi:2009hh,Wang:2009azb,Kobayashi:2010eh},
the matter bounce
\cite{Brandenberger:2009yt,Brandenberger:2009ic,Cai:2009in,Gao:2009wn,
Saridakis:2011pk},
the black hole properties
\cite{Danielsson:2009gi,Cai:2009pe,Kehagias:2009is,Myung:2009va,Park:2009zra,BottaCantcheff:2009mp,Lee:2009rm,Kiritsis:2009rx,Greenwald:2009kp},
the dark energy phenomenology
\cite{Saridakis:2009bv,Park:2009zr,Chaichian:2010yi,Jamil:2010vr},
the observational constraints on the parameters of the theory
\cite{Dutta:2009jn,Dutta:2010jh}, the astrophysical phenomenology
\cite{Kim:2009dq,Harko:2009qr,Iorio:2009qx,Iorio:2009ek,Izumi:2009ry,
Lobo:2010hv,Cardone:2010tb,Saridakis:2011eq},
the thermodynamic properties
\cite{Wang:2009rw,Cai:2009qs,Cai:2009ph,Wei:2010yw,Jamil:2010di}
etc. However, despite this extended research, there are still many
ambiguities if Ho\v{r}ava-Lifshitz gravity is reliable and capable
of a successful description of the gravitational background of our
world, as well as of the cosmological behavior of the universe
\cite{Charmousis:2009tc,Li:2009bg,Sotiriou:2009bx,Bogdanos:2009uj,Koyama:2009hc,Papazoglou:2009fj,Kimpton:2010xi,Bellorin:2010je}.

Although the foundations and the possible conceptual and
phenomenological problems of Ho\v{r}ava-Lifshitz gravity and its
associated cosmology are still an open issue, it is worth
investigating different cosmological scenarios in this
gravitational background. In this regard, it would be interesting
to examine cosmological scenarios where apart from the
gravitational sector there exists a generalized Chaplygin gas
\cite{kamenshchik,Bilic:2001cg,Bento:2002ps,Gorini:2002kf,Bento:2002yx,Bento:2003we,Bento:2003dj,Bertolami:2004ic}.

The generalized Chaplygin gas (GCG) is an alternative to the
Quintessence model which had attracted great interest in recent
times. The scenario can explain the acceleration of the universe
via an exotic equation of state, which mimics  a pressureless
fluid at the early stages of evolution of the Universe, and a
cosmological constant at late times. It is therefore interesting
to consider the GCG scenario as a unified description for dark
matter and dark energy \cite{Bento:2002ps}. The background
evolution fits well the observational data \cite{Bento:2002yx,
Bento:2003we}, however the cosmological behavior is
indistinguishable from that of the $\Lambda$CDM scenario while
fitting with the structure formation data as well as with data
from Cosmic Microwave Background radiation \cite{Sn,bean}.
Additionally, the scenario is plagued by the presence of
instabilities as well as oscillations which are not observed in
the matter power spectrum.
 Although this is a serious drawback of GCG scenario
it is not the final verdict for its fate as a model for the
unified description of dark matter and dark energy. As it was
shown by Reis {\it{et al.}} in  \cite {reis:2003}, allowing for
small entropy perturbations can eliminate instabilities and
oscillations in the matter power spectrum, even in the linear
regime, for a region of the parameter space where the GCG model
behaves quite differently from the $\Lambda$CDM one. Furthermore,
as it was shown by Avelino {\it{et al.}} in \cite{Large},  in the
GCG scenario the transition from dark matter to dark energy
behavior is never smooth, and hence the linear theory which was
used by \cite{Sn, bean} to rule out GCG as a unified model, may
break down late in the matter-dominated era, even on large
cosmological scales. Therefore, nonlinear effects should be
necessarily taken into account when confronting cosmological
observations. Moreover, the addition of baryons in the GCG
scenario can also improve the behavior of the matter power
spectrum \cite{Large2}. In summary, the GCG as a unified model of
dark matter and dark energy is not completely ruled out and it
deserves further investigations. Finally, we mention that GCG in
the presence of cold dark matter as well as baryonic matter (where
GCG acts as a normal dark energy candidate) is one of the most
well-fit scenarios with cosmological observations, amongst all the
exotic models that have been considered so far \cite{davis}.

In the present work we construct the cosmology of a generalized
Chaplygin Gas in a universe governed by Ho\v{r}ava-Lifshitz
gravity. Additionally, we use observational data from Type Ia
Supernovae (SNIa) \cite{SNIadata}, Baryon Acoustic Oscillations (BAO) \cite{BAOdata} and Cosmic
Microwave Background (CMB) \cite{CMBdata}, together with the Big Bang
Nucleosynthesis (BBN) conditions, to constrain the various
parameters of the model. Furthermore, in order to be general we
perform the analysis both with and without the detailed-balance
condition of the gravitational sector.

The manuscript is organized as follows: In section \ref{model} we
present Ho\v{r}ava-Lifshitz cosmology in both its detailed-balance
and beyond-detailed-balance version. In section \ref{GCGHL} we
construct the scenario of a generalized Chaplygin Gas in
Ho\v{r}ava-Lifshitz gravitational background and we extract the
cosmological equations. In section \ref{observations} we use
observational data in order to constrain the various parameters of
the scenario. In section \ref{cosmimpl} we discuss the
cosmological implications, focusing on the evolution of the total
equation-of-state parameter of the universe and of the expansion
rate. Finally, section \ref{conclusions} is devoted to the summary
of our results.

\section{Ho\v{r}ava-Lifshitz cosmology}
\label{model}

In this section we briefly review the scenario where the
cosmological evolution is governed by Ho\v{r}ava-Lifshitz gravity
\cite{Calcagni:2009ar,Kiritsis:2009sh}. The dynamical variables
are the lapse and shift functions, $N$ and $N_i$ respectively, and
the spatial metric $g_{ij}$ (roman letters indicate spatial
indices). In terms of these fields the full metric is written as:
\begin{eqnarray}
ds^2 = - N^2 dt^2 + g_{ij} (dx^i + N^i dt ) ( dx^j + N^j dt ) ,
\end{eqnarray}
where indices are raised and lowered using $g_{ij}$. The scaling
transformation of the coordinates reads: $
 t \rightarrow l^3 t~~~{\rm and}\ \ x^i \rightarrow l x^i
$.

\subsection{Detailed Balance}

The gravitational action is decomposed into a kinetic and a
potential part as $S_g = \int dt d^3x \sqrt{g} N ({\cal L}_K+{\cal
L}_V)$. The assumption of detailed balance \cite{hor3}
  reduces the possible terms in the Lagrangian, and it allows
for a quantum inheritance principle \cite{hor2}, since the
$(D+1)$-dimensional theory acquires the renormalization properties
of the $D$-dimensional one. Under the detailed balance condition
 the full action of Ho\v{r}ava-Lifshitz gravity is given by
\begin{eqnarray}
 S_g &=&  \int dt d^3x \sqrt{g} N \left\{
\frac{2}{\kappa^2}
(K_{ij}K^{ij} - \lambda K^2) \ \ \ \ \ \ \ \ \ \ \ \ \ \ \ \ \  \right. \nonumber \\
&+&\left.\frac{\kappa^2}{2 w^4} C_{ij}C^{ij}
 -\frac{\kappa^2 \mu}{2 w^2}
\frac{\epsilon^{ijk}}{\sqrt{g}} R_{il} \nabla_j R^l_k +
\frac{\kappa^2 \mu^2}{8} R_{ij} R^{ij}
     \right. \nonumber \\
&-&\left.    \frac{\kappa^2 \mu^2}{8( 3 \lambda-1)} \left[ \frac{1
- 4 \lambda}{4} R^2 + \Lambda  R - 3 \Lambda ^2 \right] \right\},
\label{acct}
\end{eqnarray}
where
\begin{eqnarray}
K_{ij} = \frac{1}{2N} \left( {\dot{g_{ij}}} - \nabla_i N_j -
\nabla_j N_i \right)
\end{eqnarray}
is the extrinsic curvature and
\begin{eqnarray} C^{ij} \, = \, \frac{\epsilon^{ijk}}{\sqrt{g}} \nabla_k
\bigl( R^j_i - \frac{1}{4} R \delta^j_i \bigr)
\end{eqnarray}
the Cotton tensor, and the covariant derivatives are defined with
respect to the spatial metric $g_{ij}$. $\epsilon^{ijk}$ is the
totally antisymmetric unit tensor, $\lambda$ is a dimensionless
constant and the variables $\kappa$, $w$ and $\mu$ are constants
with mass dimensions $-1$, $0$ and $1$, respectively. Finally, we
mention that in action (\ref{acct}) we have already performed the
usual analytic continuation of the parameters $\mu$ and $w$ of the
original version of Ho\v{r}ava-Lifshitz gravity, since such a
procedure is required in order to obtain a realistic cosmology
\cite{Lu:2009em,Minamitsuji:2009ii,Wang:2009rw,Park:2009zra}
(although it could fatally affect the gravitational theory
itself). Therefore, in the present work $\Lambda $ is a positive
constant, which as usual is related to the cosmological constant
in the IR limit.

In order to add the matter component we follow the hydrodynamical approach (which is most suitable for
phenomenological analysis) of adding a
cosmological stress-energy tensor to the gravitational field
equations, by demanding to recover the usual general relativity
formulation in the low-energy limit
\cite{Sotiriou:2009bx,Chaichian:2010yi,Carloni:2009jc}. Thus, this
matter-tensor is a hydrodynamical approximation with $\rho_m$ and
$p_m$ (or $\rho_m$ and $w_m$) as parameters. Note that this
$\rho_m$ is the total matter energy density, that is it accounts
for both the baryonic $\rho_b$ as well as the dark matter
$\rho_{\rm dm}$. Similarly, one can additionally include the
standard-model-radiation component (corresponding to photons and
neutrinos), with the additional parameters $\rho_r$ and $p_r$ (or
$\rho_r$ and $w_r$).

In order to investigate cosmological frameworks, we impose the
projectability condition \cite{Charmousis:2009tc} and we use an
FRW metric
\begin{eqnarray}
N=1~,~~g_{ij}=a^2(t)\gamma_{ij}~,~~N^i=0~,
\end{eqnarray}
with
\begin{eqnarray}
\gamma_{ij}dx^idx^j=\frac{dr^2}{1- K r^2}+r^2d\Omega_2^2~,
\end{eqnarray}
where $ K<,=,> 0$ corresponding  to open, flat, and closed
universe respectively (we have adopted the convention of taking
the scale factor $a(t)$ to be dimensionless and the curvature
constant $ K$ to have mass dimension 2). By varying $N$ and
$g_{ij}$, we extract the Friedmann equations:
\begin{eqnarray}\label{Fr1fluid}
H^2 &=&
\frac{\kappa^2}{6(3\la-1)}\Big(\rho_m+\rho_r\Big)\nonumber\\
&+&\frac{\kappa^2}{6(3\la-1)}\left[ \frac{3\kappa^2\mu^2
K^2}{8(3\lambda-1)a^4} +\frac{3\kappa^2\mu^2\Lambda
^2}{8(3\lambda-1)}
 \right]\nonumber\\
 &-&\frac{\kappa^4\mu^2\Lambda  K}{8(3\lambda-1)^2a^2} \ ,
\end{eqnarray}
\begin{eqnarray}\label{Fr2fluid}
\dot{H}+\frac{3}{2}H^2 &=&
-\frac{\kappa^2}{4(3\la-1)}\Big(w_m\rho_m+w_r\rho_r\Big)\nonumber\\
&-&\frac{\kappa^2}{4(3\la-1)}\left[\frac{\kappa^2\mu^2
K^2}{8(3\lambda-1)a^4} -\frac{3\kappa^2\mu^2\Lambda
^2}{8(3\lambda-1)}
 \right]\nonumber\\
 &-&\frac{\kappa^4\mu^2\Lambda  K}{16(3\lambda-1)^2a^2}\ ,
\end{eqnarray}
where  $H\equiv\frac{\dot a}{a}$ is the Hubble parameter.

The term proportional to $a^{-4}$ is the usual ``dark radiation
term'', present in Ho\v{r}ava-Lifshitz cosmology
\cite{Calcagni:2009ar,Kiritsis:2009sh}, while the constant term is
just the explicit cosmological constant. Finally, as usual,
$\rho_m$ follows the standard evolution equation
\begin{eqnarray}\label{rhodotfluid}
&&\dot{\rho}_m+3H(\rho_m+p_m)=0,
\end{eqnarray}
while $\rho_r$ (standard-model radiation) follows
\begin{eqnarray}\label{rhodotfluidrad}
&&\dot{\rho}_r+3H(\rho_r+p_r)=0.
\end{eqnarray}

If we require expressions (\ref{Fr1fluid}) to coincide with the
standard Friedmann equations, in units where $c=1$  we set
\cite{Calcagni:2009ar,Kiritsis:2009sh}:
\begin{eqnarray}
G_{\rm cosmo}&=&\frac{\kappa^2}{16\pi(3\lambda-1)}\nonumber\\
\frac{\kappa^4\mu^2\Lambda}{8(3\lambda-1)^2}&=&1,
\label{simpleconstants0}
\end{eqnarray}
where $G_{\rm cosmo}$ is the ``cosmological'' Newton's constant,
that is the one that is read from the Friedmann equations. We
mention that in theories with Lorentz invariance breaking $G_{\rm
cosmo}$ does not coincide with the
 ``gravitational'' Newton's constant
$G_{\rm grav}$, that is the one that is read from the action,
unless Lorentz invariance is restored \cite{Carroll:2004ai}. For
completeness we mention that in our case
\begin{eqnarray}
G_{\rm grav}=\frac{\kappa^2}{32\pi}\label{Ggrav},
\end{eqnarray}
as it can be straightforwardly read from the action (\ref{acct})
(our definitions of $G_{\rm cosmo}$, $G_{\rm grav}$ coincide with
those of \cite{Blas:2009qj,Papazoglou:2009fj}). Thus, it becomes
obvious that in the IR ($\lambda=1$), where Lorentz invariance is
restored, $G_{\rm cosmo}$ and $G_{\rm grav}$ coincide.

In our work the running of $\lambda$ is not a problem, since the
whole relevant cosmological history, that is after inflation, is
obviously inside the IR limit, that is with $\lambda=1$. On the
other hand, the divergence of $\lambda$ from $1$, that is the
quantum gravitational features of the theory, become significant
at very early times, that is close to the Big Bang or the
cosmological bounce. This was additionally shown in
\cite{Dutta:2010jh}, where a detailed analysis based on SNIa, BAO
and CMB observations, as well as BBN considerations, constrained
$\lambda$ inside a narrow window around its infrared value:
$|\lambda-1|<0.02$. In summary, since in this work we are
interested in post-inflation evolution, we set $\lambda=1$, and
thus we simplify our notation using $G$ for the  (coincided
gravitational and cosmological) Newton's constant.

\subsection{Beyond Detailed Balance}

The aforementioned formulation of Ho\v{r}ava-Lifshitz cosmology
has been performed under the imposition of the detailed-balance
condition. However, in the literature there is a discussion
whether this condition leads to reliable results or if it is able
to reveal the full information of Ho\v{r}ava-Lifshitz
 gravity \cite{Calcagni:2009ar,Kiritsis:2009sh}. Therefore, one
 needs to investigate also the Friedman equations in the case
 where detailed balance is relaxed. In such a case one can in
 general write
 \cite{Charmousis:2009tc,Sotiriou:2009bx,Bogdanos:2009uj,Carloni:2009jc,Leon:2009rc}:
\begin{eqnarray}\label{Fr1c}
H^2 &=&
\frac{2\sigma_0}{(3\la-1)}\Big(\rho_m+\rho_r\Big)\nonumber\\
&+&\frac{2}{(3\la-1)}\left[ \frac{\sigma_1}{6}+\frac{\sigma_3
K^2}{6a^4} +\frac{\sigma_4 K}{6a^6}
 \right]\nonumber\\&+&\frac{\sigma_2}{3(3\la-1)}\frac{ K}{a^2}
\end{eqnarray}
\begin{eqnarray}\label{Fr2c}
\dot{H}+\frac{3}{2}H^2 &=&
-\frac{3\sigma_0}{(3\la-1)}\Big(w_m\rho_m+w_r\rho_r\Big)\nonumber\\
&-&\frac{3}{(3\la-1)}\left[ -\frac{\sigma_1}{6}+\frac{\sigma_3
K^2}{18a^4} +\frac{\sigma_4 K}{6a^6}
 \right]\nonumber\\&+&
 \frac{\sigma_2}{6(3\la-1)}\frac{ K}{a^2},
\end{eqnarray}
where $\sigma_0\equiv \kappa^2/12$, and the constants $\sigma_i$
are arbitrary (with $\sigma_2$ being negative and $\sigma_4$
positive). As we observe, the effect of the detailed-balance
relaxation is the decoupling of the coefficients, together with
the appearance of a term proportional to $a^{-6}$.

Finally, if we force (\ref{Fr1c}),(\ref{Fr2c}) to coincide with
 the standard Friedmann equations, we obtain:
\begin{eqnarray}
&&G_{\rm cosmo}=\frac{6\sigma_0}{8\pi(3\lambda-1)}\nonumber\\
&&\sigma_2=-3(3\lambda-1), \label{simpleconstants0nd}
\end{eqnarray}
while in this case the ``gravitational'' Newton's constant $G_{\rm
grav}$ reads \cite{Sotiriou:2009bx}:
\begin{eqnarray}
G_{\rm grav}=\frac{6\sigma_0}{16\pi}\label{Ggravbdb}.
\end{eqnarray}
Similarly to the detailed balance case, in the IR ($\lambda=1$)
$G_{\rm cosmo}$ and $G_{\rm grav}$ coincide, and thus in the
following we will use $G$ to denote the  (coincided gravitational
and cosmological) Newton's constant.

\section{Generalized Chaplygin gas in Ho\v{r}ava-Lifshitz cosmology}
\label{GCGHL}

As discussed in the introduction, Chaplygin gas scenarios have been studied extensively as possible
candidates for a unified description of dark matter and dark
energy through a single fluid. The generalized
Chaplygin gas is a fluid with an equation of state given by:
\begin{equation}
\label{rhochaplygin} p_c=-\frac{A}{\rho_c^\beta},
\end{equation}
with $A$ a positive constant and $0<\beta\leq1$. $\rho_c$ and
$p_c$ denote the energy density and pressure of the Chaplygin gas,
and henceforth a subscript ``c'' will denote quantities pertaining
to the Chaplygin gas. The simple Chaplygin gas is the special case
of $\beta=1$. Finally, considering the covariant conservation of
energy-momentum, $\rho_c$ satisfies the standard evolution
equation
\begin{eqnarray}\label{rhodotfluid}
&&\dot{\rho}_c+3H(\rho_c+p_c)=0.
\end{eqnarray}

Equations (\ref{rhochaplygin}) and (\ref{rhodotfluid}) can be
easily solved to yield \cite{Bilic:2001cg,Bento:2002ps}:
\begin{eqnarray}
\label{rhochaplsol}
 \rho_c=\rho_{c0}\[A_s+\frac{1-A_s}{a^{3\(1+\beta\)}}\]^{\frac{1}{1+\beta}}
\end{eqnarray}
where $\rho_{c0}$ denotes  the present-day density of the
Chaplygin gas (in what follows, a subscript 0 will always denote
the present-day value of a quantity), and we have introduced the
parameter
 \begin{eqnarray}
 A_s\equiv
A\rho_{c0}^{-(1+\beta)} .
\end{eqnarray}
Thus, (\ref{rhochaplygin}) provides the pressure of the Chaplygin
gas as
\begin{eqnarray}
\label{pchaplsol}
p_c=-\rho_{c0}A_s\[A_s+\frac{1-A_s}{a^{3\(1+\beta\)}}\]^{-\frac{\beta}{1+\beta}}.
\end{eqnarray}
Clearly, $A_s$ is therefore just the present value of the
equation-of-state parameter of the generalized Chaplygin gas.
Furthermore, $\beta$ is related to the sound velocity of the
Chaplygin gas at the present time, which is given by $\beta A_s$
\cite{Bento:2003we}.

The most important advantage of the Chaplygin gas scenario is that
it behaves like dark matter at early times ($a\ll1$) and as a
cosmological constant at late times ($a\gg1$), smoothly
interpolating between the two phases. Thus, since in this work we
are interested in investigating the generalized Chaplygin gas in
Ho\v{r}ava-Lifshitz cosmology, we first separate the total matter
energy density $\rho_m$, that was present in the Friedmann
equations of the previous section, into the baryonic $\rho_b$ and
the dark matter $\rho_{\rm dm}$ parts, and then we use the
generalized Chaplygin gas energy density $\rho_c$ instead of
$\rho_{\rm dm}$. Therefore, using also the definitions
(\ref{simpleconstants0}), the Friedmann equations
\eqref{Fr1fluid},\eqref{Fr2fluid} of the detailed balance case
take the form:
\begin{equation}\label{Fr1finalb}
H^2 = \frac{8\pi G}{3}\left[\rho_b+ \rho_c+\rho_r \right]+\left[
\frac{K^2}{2\Lambda a^4}+\frac{\Lambda}{2} \right]-\frac{K}{a^2}
\end{equation}
\begin{equation}\label{Fr2finalb}
\dot{H}+\frac{3}{2}H^2 = -4\pi G
\left[p_c+\frac13\rho_r\right]-\left[\frac{K^2}{4\Lambda
a^4}-\frac{3\Lambda}{4}\right]-\frac{K}{2a^2}.
\end{equation}
Similarly, for the beyond-detailed-balance case the Friedmann
equations (\ref{Fr1c}),(\ref{Fr2c}), using also the
identifications (\ref{simpleconstants0nd}), become
\begin{align}
 \label{Fr1finalbbdb}
H^2 =& \frac{8\pi
G}{3}\left[\rho_b+ \rho_c+\rho_r \right]\nonumber\\
&+\frac{2}{(3\la-1)}\left[ \frac{\sigma_1}{6}+\frac{\sigma_3
K^2}{6a^4} +\frac{\sigma_4 K}{6a^6}
 \right]-\frac{K}{a^2}
\end{align}
\begin{eqnarray}\label{Fr2finalbbdb}
&&\dot{H}+\frac{3}{2}H^2 = -4\pi G
\left[p_c+\frac13\rho_r\right]\nonumber\\
&&\ \ \ \ \ \ \ -\frac{3}{(3\la-1)}\left[
-\frac{\sigma_1}{6}+\frac{\sigma_3 K^2}{18a^4} +\frac{\sigma_4
K}{6a^6}
 \right]-\frac{K}{2a^2}.\ \ \
\end{eqnarray}

\section{Observational Constraints}
\label{observations}

In the conventional gravitational background of general
relativity, the cosmological scenarios of generalized Chaplygin
gas have been extensively constrained using observational data. In
particular, in \cite{Bertolami:2004ic,Gravsn,Large,Sn,Sn2} SNIa
observations were used, in \cite{Large,Large2} data from large
scale structure, in \cite{Cmb,Cmb2} CMB data, while in
\cite{Gravsn} the authors used  gravitational lensing
observations. Finally, constraints from combined data sources have
been obtained in \cite{Park:2009np} and \cite{delCampo:2009cz}.

In the present work we will consider observations in order to
constraint the scenario of a generalized Chaplygin gas in a
universe governed by Ho\v{r}ava-Lifshitz gravity. In particular we
will use data from SNIa, BAO, CMB and BBN  to constrain the Chaplygin gas
parameters together with the parameters of Ho\v{r}ava-Lifshitz
gravity. For completeness, we perform our analysis
separately for the detailed-balance and the beyond-detailed-balance
scenario.

\subsection{Constraints on the Detailed-Balance scenario}

We first consider the detailed-balance version of the theory. It
proves more convenient to use the redshift $z$ instead of the
scale factor as the independent variable, through the relation
$1+z\equiv a_0/a=1/a$. Furthermore, we use the
dimensionless density parameters
\begin{eqnarray}
\label{densparam} \Omega_i\equiv \frac{8\pi G}{3H^2}\rho_i,\ \
\Omega_ K\equiv - \frac{K}{H^2a^2},
\end{eqnarray}
and we introduce the dimensionless parameter
\begin{eqnarray}
\label{densparam2} \omega\equiv \frac{\Lambda}{2 H_0^2}.
\end{eqnarray}
Finally, we use the
dimensionless expansion rate
\begin{eqnarray}
\label{fidexprate}
  E(z)\equiv \frac{H(z)}{H_0}.
\end{eqnarray}

Using the above definitions, the Friedman equation
(\ref{Fr1finalb}) becomes:
\begin{align}
E^2(z)=&\,\Omega_{b0}(1+z)^3+\Omega_{c0}F(z)+\Omega_{r0}(1+z)^4\nonumber\\
&+\,\Omega_{ K0}(1+z)^2+\Big[\omega+\frac{\Omega_{ K
0}^2}{4\omega}(1+z)^4\Big],
 \label{Frdbfinal}
\end{align}
where \be F(z)=\left[ A_{s}+\(1-A_{s}\)(1+z)^{3(1+\beta)}\right]
^{\frac{1}{1+\beta}}.
  \ee
 Obviously, \eqref{Frdbfinal} is subject
to the present-time constraint $E(z=0)=1$, which leads to the
condition:
 \be \label{cond1}
\Omega_{b0}+\Omega_{c0}+\Omega_{r0}+\Omega_{K0}+\omega+\frac{\Omega_{K0}^2}{4\omega}=1.
\ee

\begin{center}
\begin{figure*}[ht]
\begin{tabular}{c@{\qquad}c}
\epsfig{file=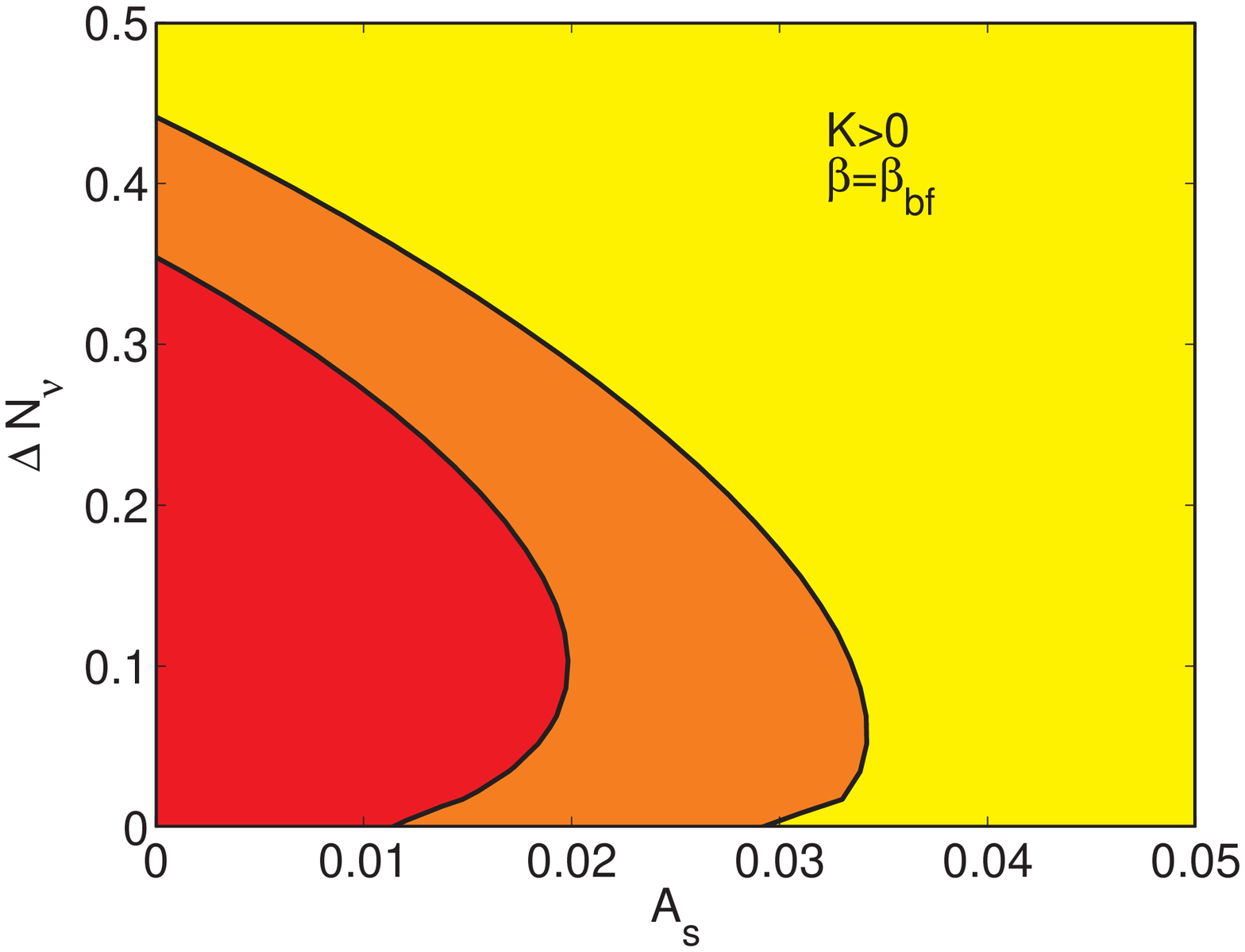,width=6 cm}&\epsfig{file=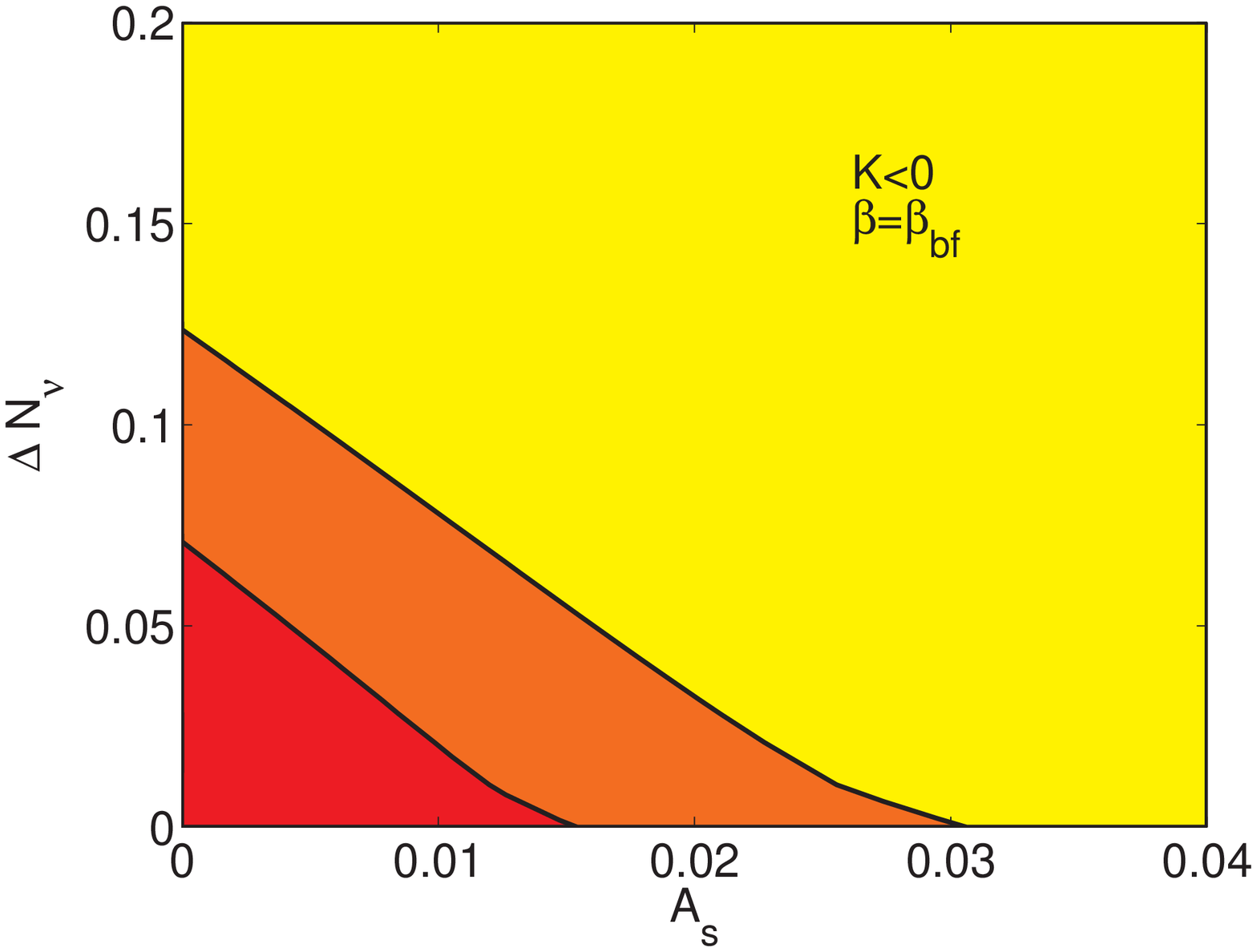,width=6 cm}\\
\epsfig{file=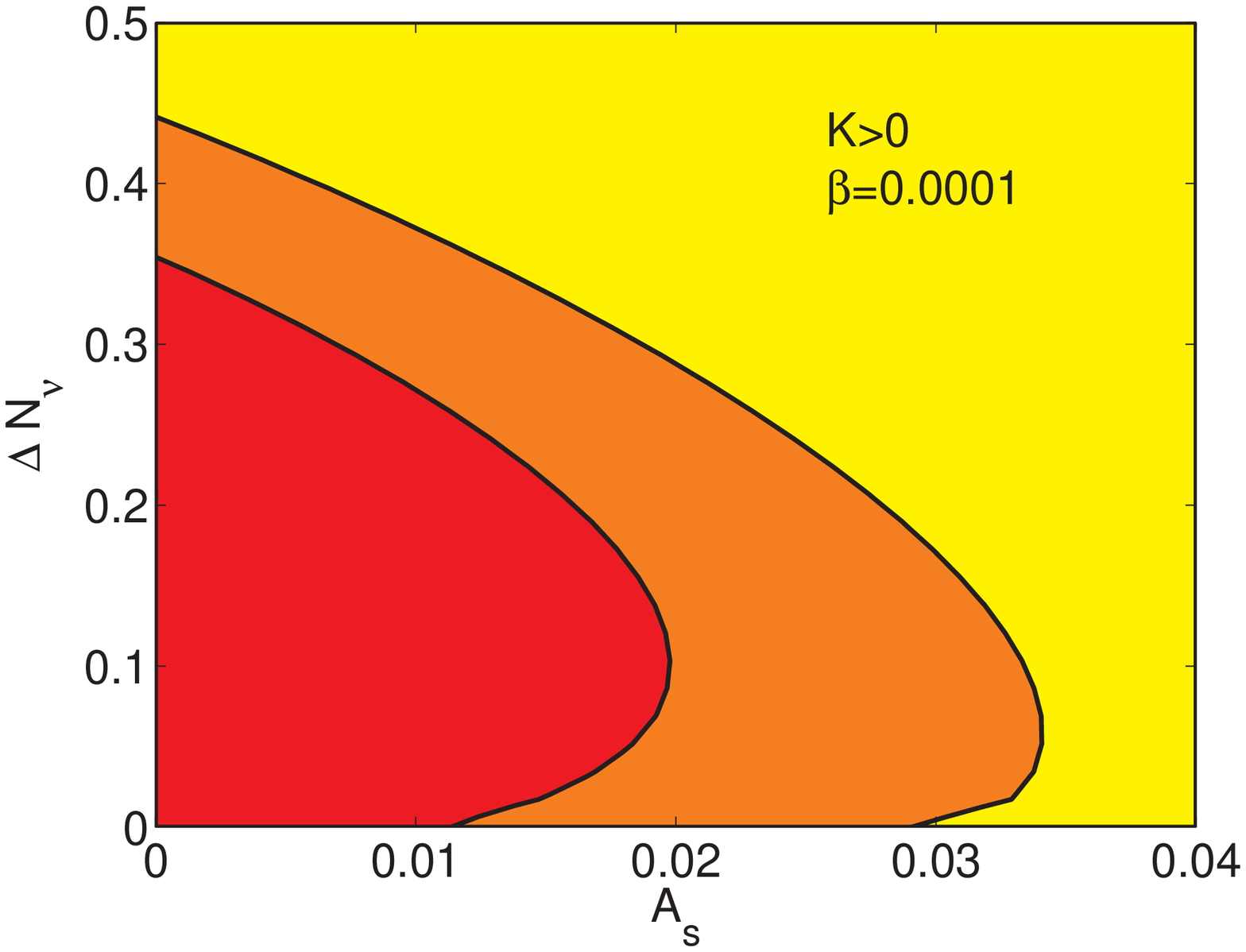,width=6 cm}&\epsfig{file=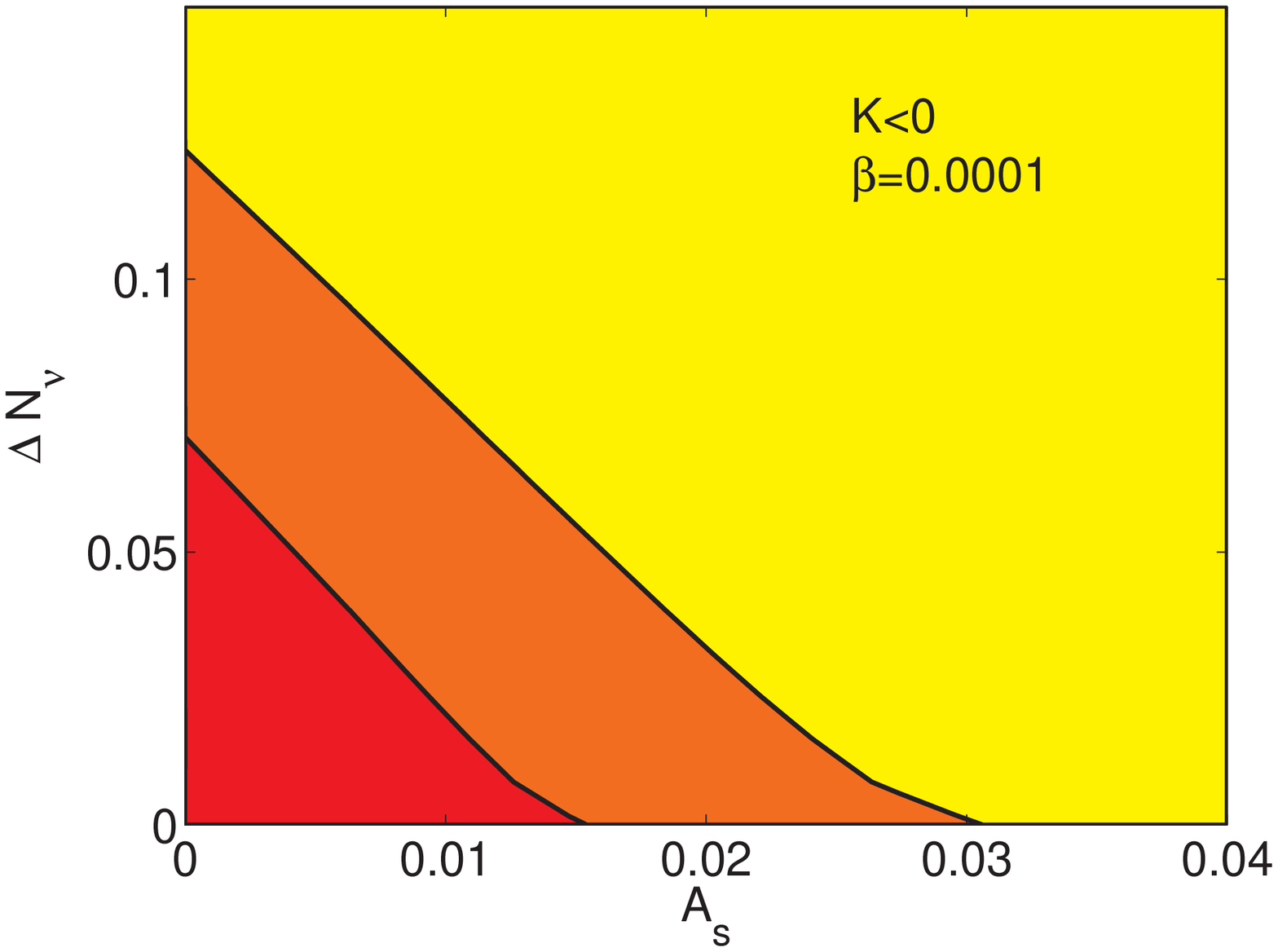,width=6 cm}\\
\epsfig{file=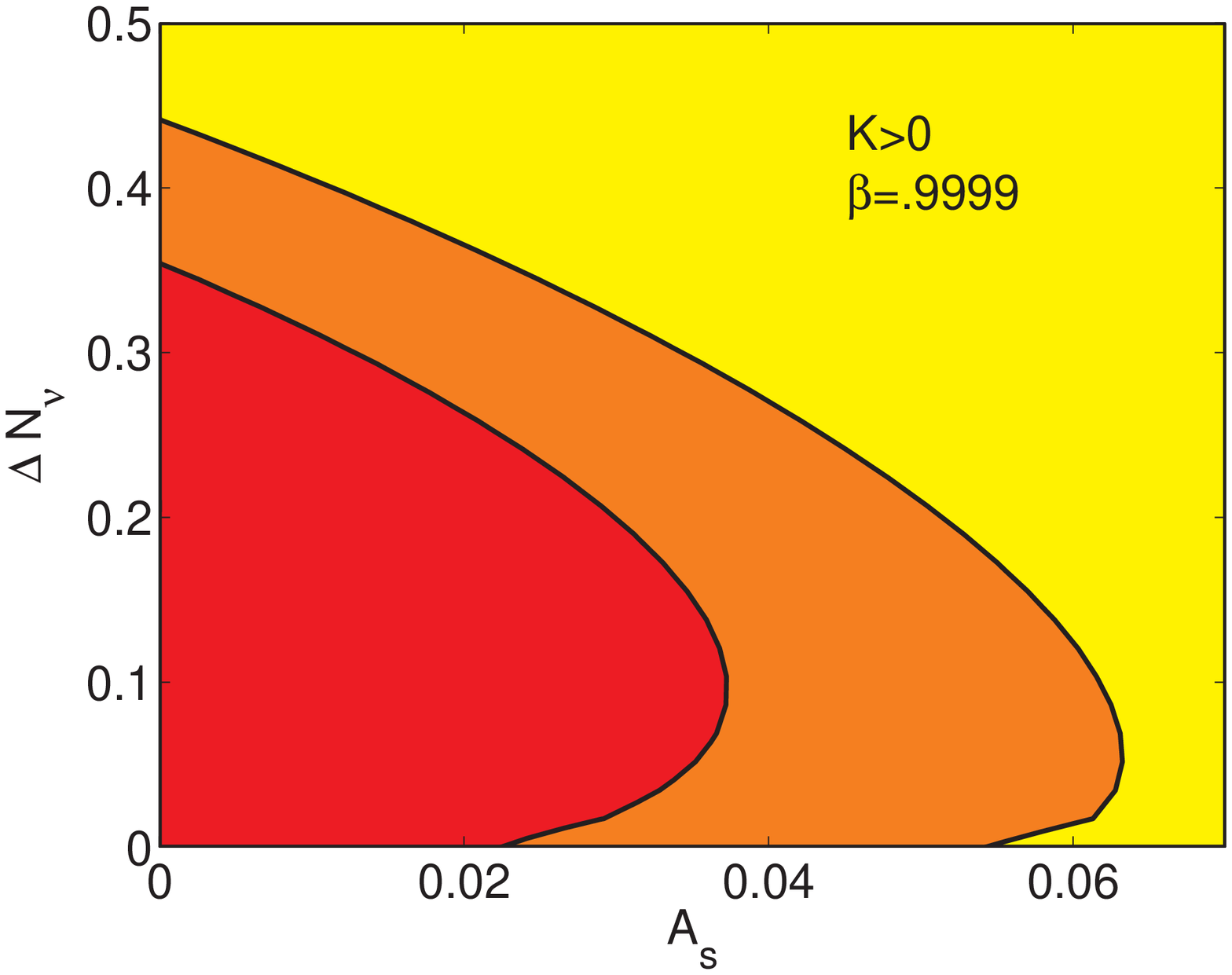,width=6 cm}&\epsfig{file=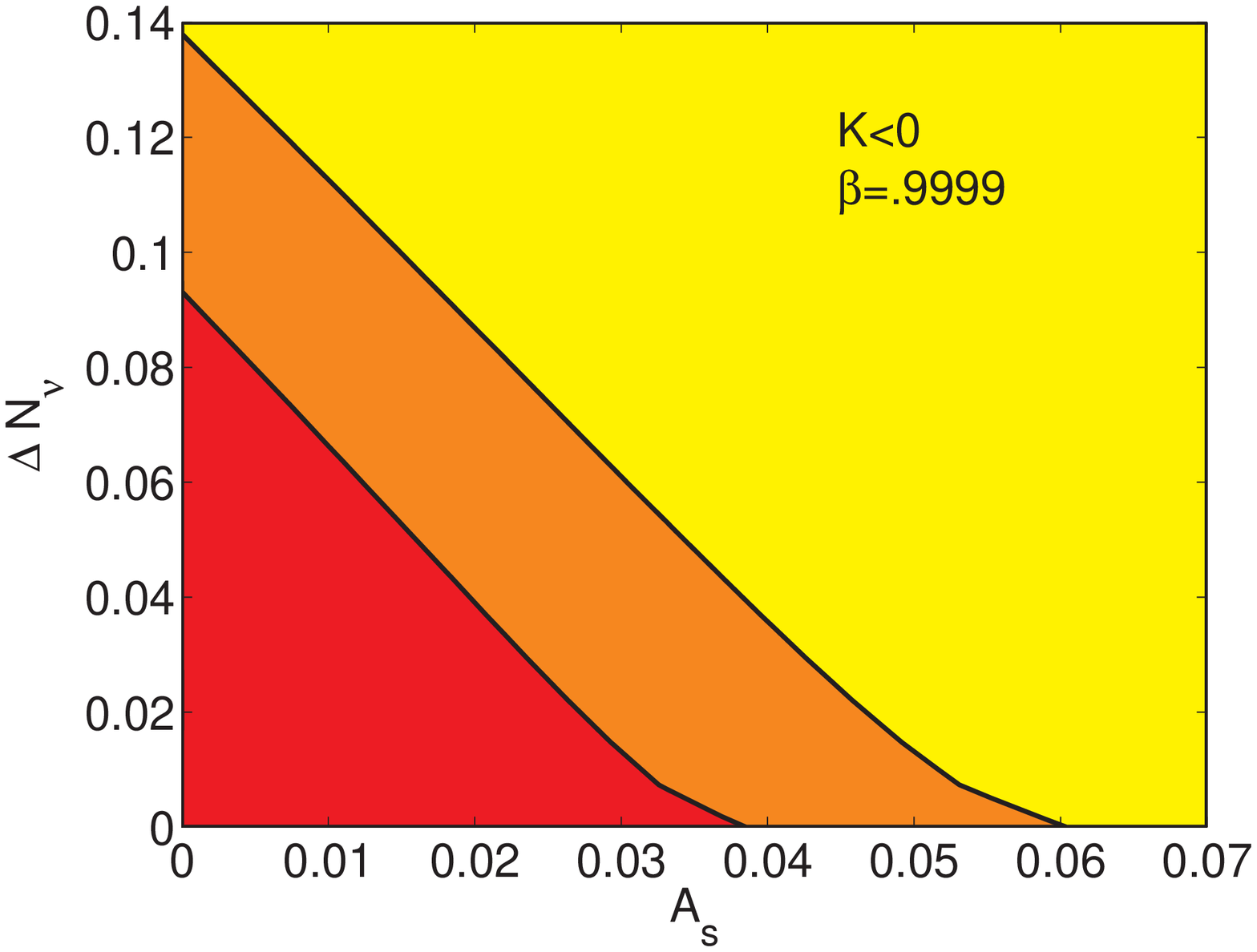,width=6 cm}
\end{tabular}
\caption{(Color online) {\it{ \textbf{Detailed Balance:}
Likelihood contours of $A_s$ and $\dn$ for different choices of
$\beta$ as discussed in the text. The  lightest region is
excluded at the 2$\sigma$ level, and the darker region is
excluded at the 1$\sigma$ level.  The darkest region is not
excluded at either confidence level.}}} \label{dbcontours}
\end{figure*}
\end{center}

As we have already mentioned above, the term $\Omega_{ K
0}^2/(4\omega)$ is the coefficient of the dark radiation term,
which is a characteristic feature of the Ho\v{r}ava-Lifshitz
gravitational background. Since this dark radiation component has
been present also during the time of nucleosynthesis, it is
subject to bounds from Big Bang Nucleosynthesis (BBN). As
discussed in detail in \cite{Dutta:2009jn},  if the upper limit on
the total amount of Ho\v{r}ava-Lifshitz dark radiation and
kination-like (a quintessence field dominated by kinetic energy
\cite{kination,kination1}) components allowed during BBN is
expressed through the parameter $\Delta N_\nu$ of the effective
neutrino species \cite{BBNrefs,BBNrefs1,BBNrefs2,Malaney:1993ah},
then we obtain the following constraint:
  \be
\label{cond2}
 \frac{\Omega_{ K 0}^2}{4\omega}=0.135\dn \Omega_{r0}.
  \ee
In this work, in order to ensure consistency with BBN, we adopt an upper limit of $\dn\leq2.0$ following \cite{BBNrefs2}.

In most studies of dark energy models it is usual to ignore
curvature (e.g.\cite{DaveCaldwellSteinhardt, LiddleScherrer,
Dutta,Dutta1,Dutta2,Dutta3,Dutta4,Dutta5,Dutta6}), especially concerning
observational constraints. This consideration is well motivated
since most inflationary scenarios predict a high degree of spatial
flatness, and moreover the CMB data impose stringent constraints
on spatial flatness in the context of constant-$w$ models (for
example a combination of WMAP+BAO+SNIa data \cite{Komatsu:2008hk}
provides the tight simultaneous constraints
$-0.0179\leq\Omega_{K0}\leq0.0081$ and $-0.12\leq1+w\leq0.14$,
both at 95\% confidence).

However, it is important to keep in mind that due to degeneracies
in the CMB power spectrum (see \cite{crooks} and references
therein), the limits on curvature depend on assumptions regarding
the underlying dark energy scenario. For example, if we use a
linear $w$ (that is $w\(a\)=w_0+\(1-a\)w_a$) instead of a constant
one, the error on $\Omega_{K0}$ is of the order of a few percent,
that is much larger \cite{Wang2,Verde,Ichi1} (see
\cite{Wright,Ichi2,Ichi3} for the constraints on curvature for
different parameterizations). Additionally, in \cite{Ichi3} it was
shown that for some models of dark energy the constraint on the
curvature is at the level of $5\%$ around a flat universe, whereas
for others the data are consistent with an open universe with
$\Omega_{K0}\sim0.2$. In \cite{Verde} it was proposed that
geometrical tests, such as the combination of the Hubble parameter
$H(z)$ and the angular diameter distance $D_A(z)$, using (future)
data up to sufficiently high redshifts $z\sim 2$, might be able to
disentangle curvature from dark energy evolution, though not in a
model-independent way. Furthermore, in \cite{Cortes,Virey} the
authors highlighted the pitfalls arising from ignoring curvature
in studies of dynamical dark energy, and recommended to treat
$\Omega_{K0}$ as a free parameter to be fitted along with the
other model parameters. Lastly, note that in the present work the
spatial curvature plays a very crucial role, since
Ho\v{r}ava-Lifshitz cosmology coincides completely with
$\Lambda$CDM if one ignores curvature
\cite{Calcagni:2009ar,Kiritsis:2009sh}. Therefore, and following
the discussion above, we prefer to treat $\Omega_{K0}$ as a free
parameter.

In summary, the scenario  at hand involves the following
parameters: $\Omega_{b0}$, $\Omega_{r0}$, $\Omega_{c0}$,
$\Omega_{k0}$, $\omega$, $\dn$, $H_0$, $A_s$ and $\beta$. We fix
the parameters $\Omega_{m0}(\equiv\Omega_{b0}+\Omega_{c0})$,
$\Omega_{b0}$, $H_0$ and $\Omega_{r0}$ at their 7-year WMAP
best-fit values \cite{Komatsu:2010fb}, namely $\Omega_{m0}=0.27$,
$\Omega_{r0}=8.14\times 10^{-5}$ and $H_0=$71.4 Km/sec/Mpc.
Therefore, there are five remaining parameters, namely
$\Omega_{k0}$, $\omega$, $\dn$, $A_s$ and $\beta$, which are
subject to the two constraint  equations (\ref{cond1}) and
(\ref{cond2}). We choose to treat $\dn$, $A_s$ and $\beta$ as free
parameters and use these constraints to eliminate $\Omega_{K0}$
and $\omega$ for a given choice of curvature, as: \ba
     \label{invert1}
\omega\(K;\dn, A_s,\beta\)=1-\Omega_{m0}-(1-.135\dn)\Omega_{r0}\nonumber&&\\
-0.73\text{
sgn}\(K\)\sqrt{\dn}\sqrt{\Omega_{r0}-\Omega_{m0}\Omega_{r0}-\Omega_{r0}^2}&&
\ea
  and
      \be \label{invert2}
\vert\Omega_{K0}\(\dn,A_s,\beta\)\vert=\sqrt{0.54\,\dn\,\Omega_{r0}\,\omega\(K;\dn,
A_s,\beta\)}. \ee
 Note that in order to obtain $\omega$ and
$\Omega_{K0}$ from the above equations \eqref{invert1} and
\eqref{invert2}, the type of curvature $K$ has to be chosen a
priori. Therefore, we treat positive ($K>0$) and negative ($K<0$)
curvature separately. Finally, note that according to
\eqref{invert2} if $\dn=0$ then it forces the spatial curvature to
be zero.

Let us now proceed to constrain the three free parameters $\dn$,
$A_s$ and $\beta$. We perform a likelihood analysis using the
SNIa, BAO and CMB data (see the appendix of \cite{Dutta:2009jn}
for the details of the procedure). In Fig. \ref{dbcontours} we
present the likelihood contours of $A_s$ and $\dn$ for three
different choices of $\beta$, one at its best-fit value, and the
other two at the extremes of its allowed values. As we observe,
the scenario at hand is in general compatible with observations.
Furthermore, note that the contours are smaller for negative
curvature, indicating that negative curvature is generally
disfavored for these models. Finally, the best-fit values together
with the corresponding $1\sigma$ bounds arising from the analysis,
for the two choices of curvature, are shown in Table
\ref{db_bestfits}.
\begin{table}[ht]
    \centering
        \begin{tabular}{|c|c|c|}
        \hline
        ~& \textbf{$K<0$}        & \textbf{$K>0$}\\\hline
        $A_s$           & $1.96\times10^{-13};\,\(0,0.04\)$ &  $3.75\times10^{-11};\,\(0,0.04\)$                      \\\hline
        $\dn$             &  $0;\,\(0,0.09\)$  &     $2.75\times10^{-2};\,\(0,0.09\)$ \\\hline
        $\beta$             & $3.21\times10^{-5};\,\(0,1\)$               &       $3.89\times10^{-3};\,\(0,1\)$    \\\hline
        \end{tabular}
    \caption{Best-fit values and $1\sigma$ confidence intervals for the free parameters $A_s$, $\dn$ and $\beta$
    in the detailed balance scenario, arising from a likelihood analysis using SNIa, BAO and CMB data.}
    \label{db_bestfits}
\end{table}

\subsection{Constraints on the Beyond-Detailed-Balance scenario}

In this subsection we consider the cosmology of a generalized
Chaplygin gas in Ho\v{r}ava-Lifshitz gravity where the detailed
balance condition has been relaxed. Our cosmic fluid consists of
the baryonic matter, the standard-model radiation and the
Chaplygin gas. Starting with the Friedmann equation \eqref{Fr1c}
and proceeding similarly to the previous subsection, we can write
the dimensionless Hubble parameter $E(z)$ as:
\begin{align}
E^2(z)=&\,\Omega_{b0}(1+z)^3+\Omega_{c0}F(z)\nonumber\\
+&\,\Omega_{r0}(1+z)^4+\Omega_{ K0}(1+z)^2\nonumber\\
+&\,\Big[\omega_1+\omega_3(1+z)^4+\omega_4(1+z)^6\Big],
 \label{Frndbfinal}
\end{align}
where as usual \be F(z)=\left[
A_{s}+\(1-A_{s}\)(1+z)^{3(1+\beta)}\right] ^{\frac{1}{1+\beta}}.
\ee
 In (\ref{Frndbfinal}) we have introduced the
dimensionless parameters $\w_1$, $\w_3$ and $\w_4$, related to the
model parameters $\sigma_1$, $\sigma_3$ and $\sigma_4$
 through:
\ba
\omega_1&=&\frac{\sigma_1}{6H_0^2}\nonumber\\
\omega_3&=&\frac{\sigma_3 H_0^2\Omega_{ K0}^2}{6}\nonumber\\
\omega_4&=&-\frac{\sigma_4\Omega_{ K0}}{6}.
  \ea
   Furthermore, we consider the
combination $\omega_4$ to be positive, in order to ensure that the
Hubble parameter is real for all redshifts (note that $\omega_4>0$
is required also for the stability of the gravitational
perturbations of the theory
\cite{Sotiriou:2009bx,Bogdanos:2009uj}). For convenience we
moreover assume $\sigma_3\geq0$, that is $\omega_3\geq0$.

\begin{center}
\begin{figure*}[!]
\begin{tabular}{c@{\qquad}c}
\epsfig{file=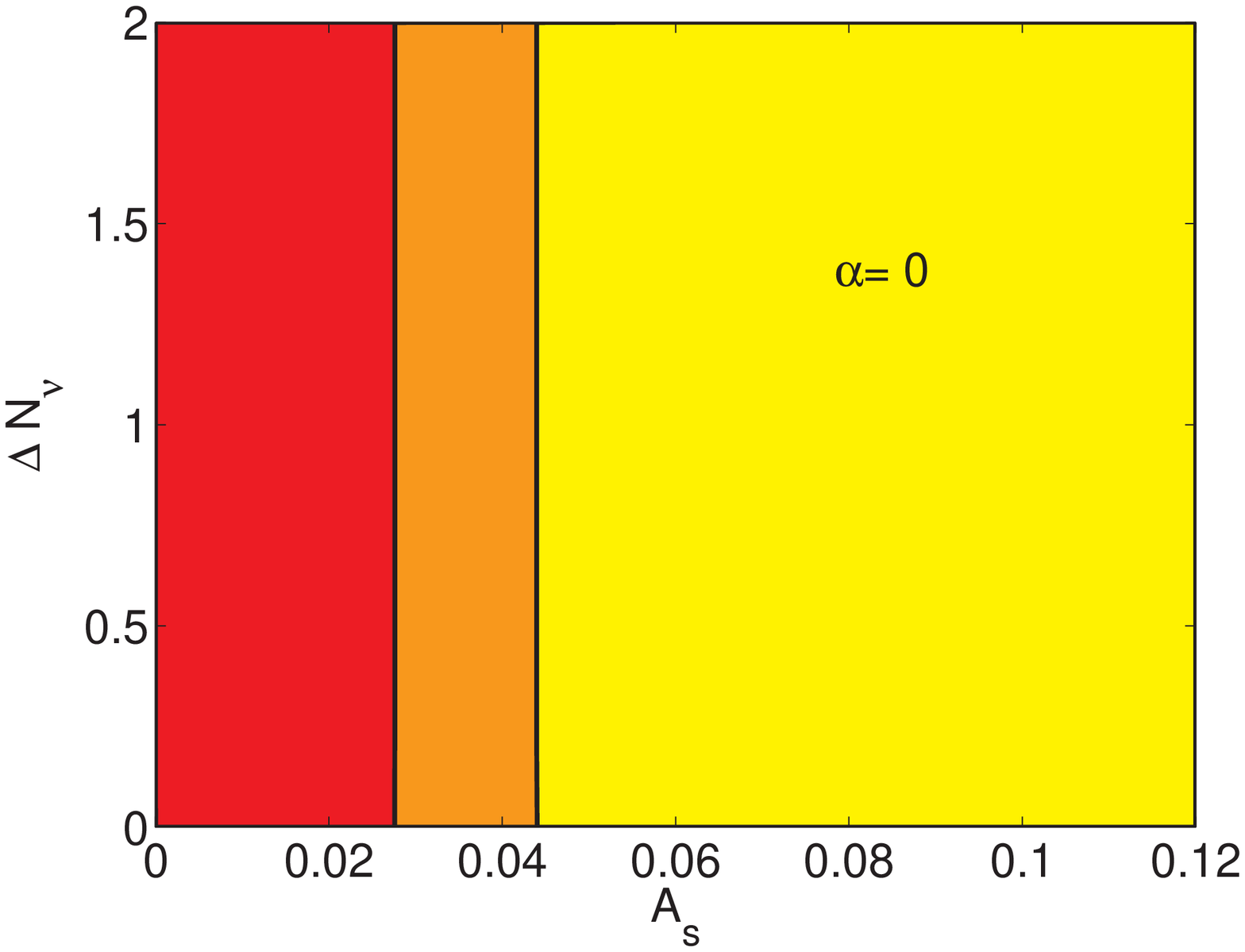,width=6 cm}&\epsfig{file=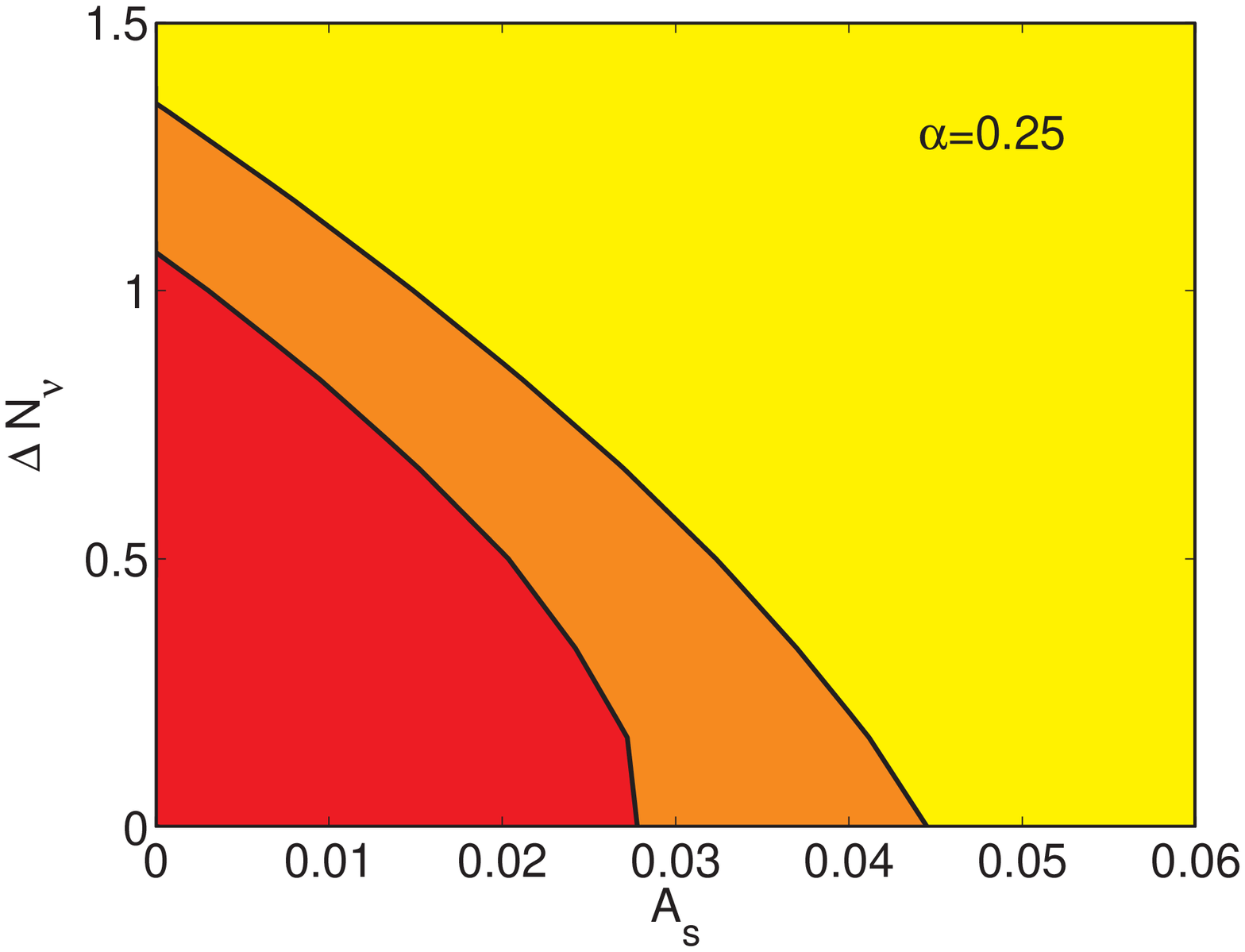,width=6 cm}\\
\epsfig{file=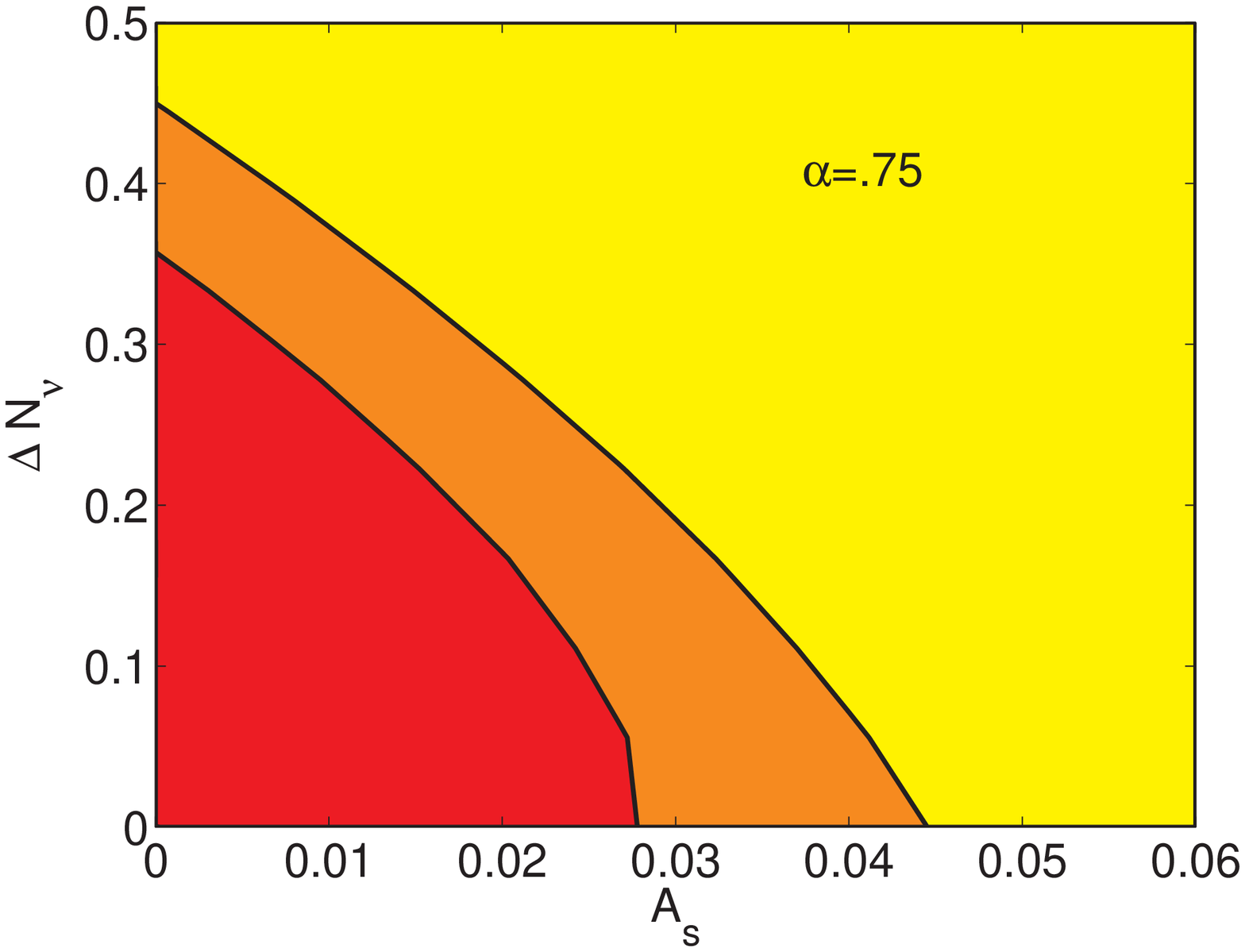,width=6
cm}&\epsfig{file=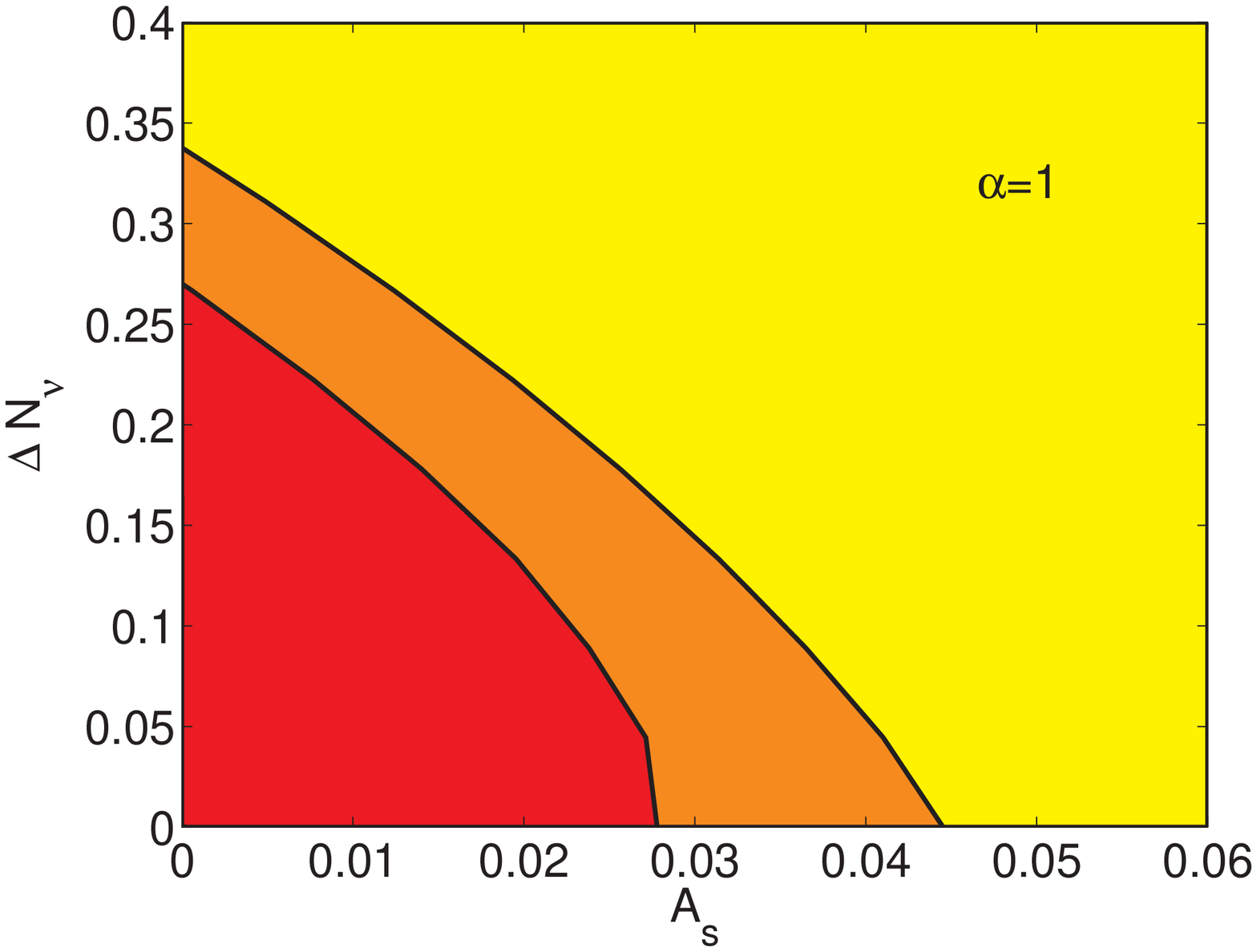,width=6 cm}
\end{tabular}
\caption{(Color online) {\it{\textbf{Beyond Detailed Balance:}
Likelihood contours of $A_s$ and $\dn$ for four choices of
$\alpha$, with the other parameters fixed at their best-fit
values. Color scheme as in Fig. \ref{dbcontours}}.}}
\label{cont_varying_alpha}
\end{figure*}
\end{center}
\begin{center}
\begin{figure*}[!]
\begin{tabular}{c@{\qquad}c}
\epsfig{file=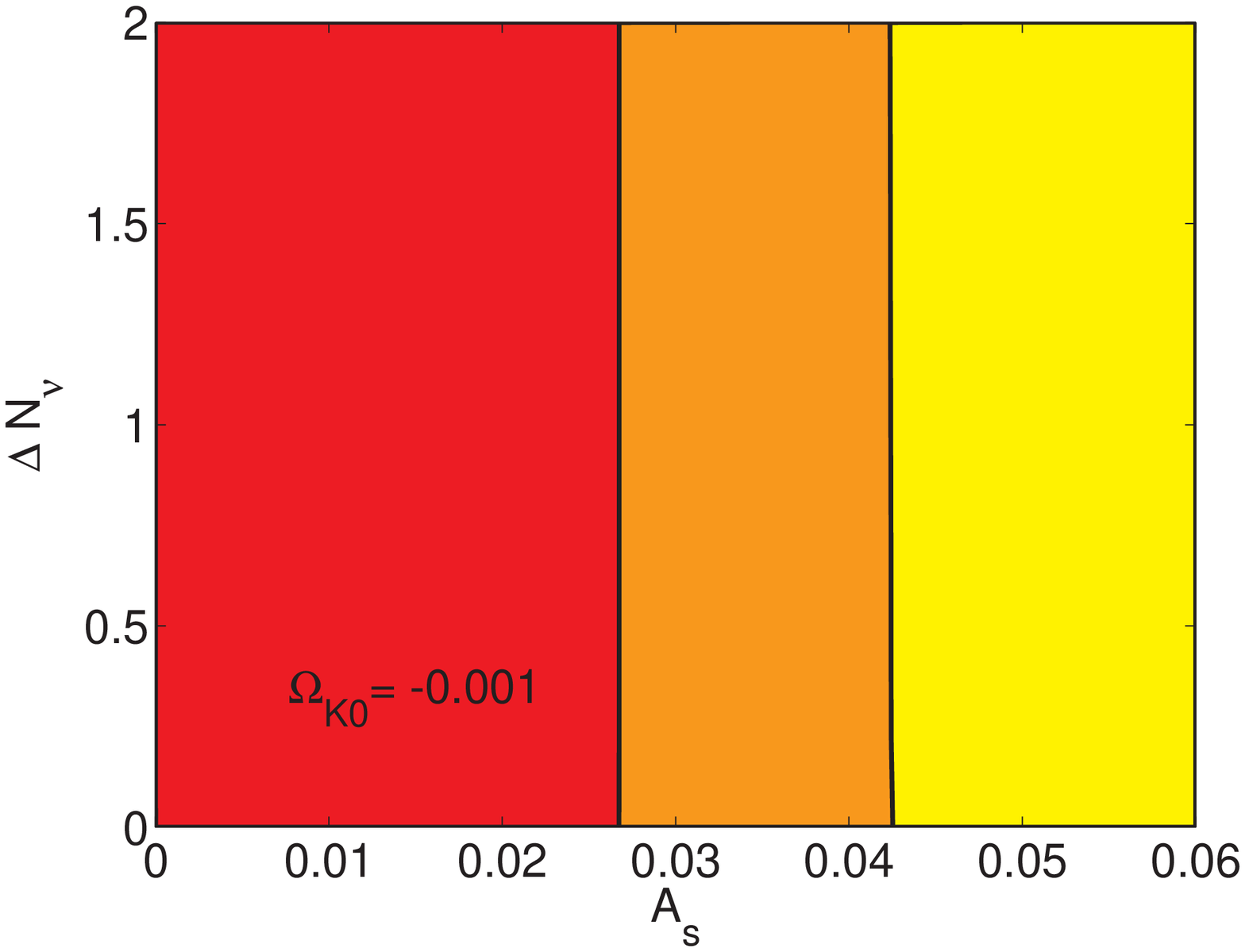,width=6 cm}&\epsfig{file=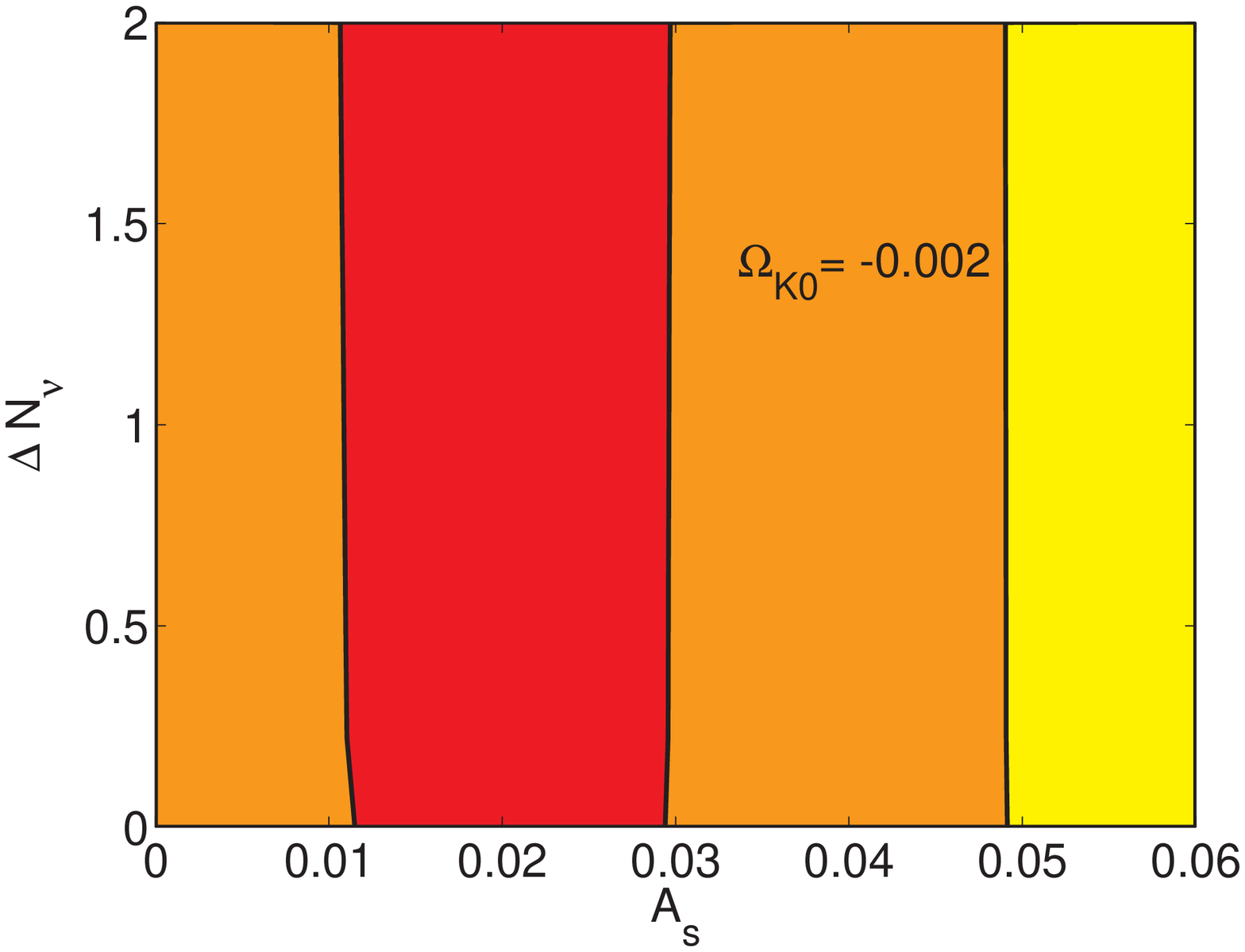,width=6 cm}\\
\epsfig{file=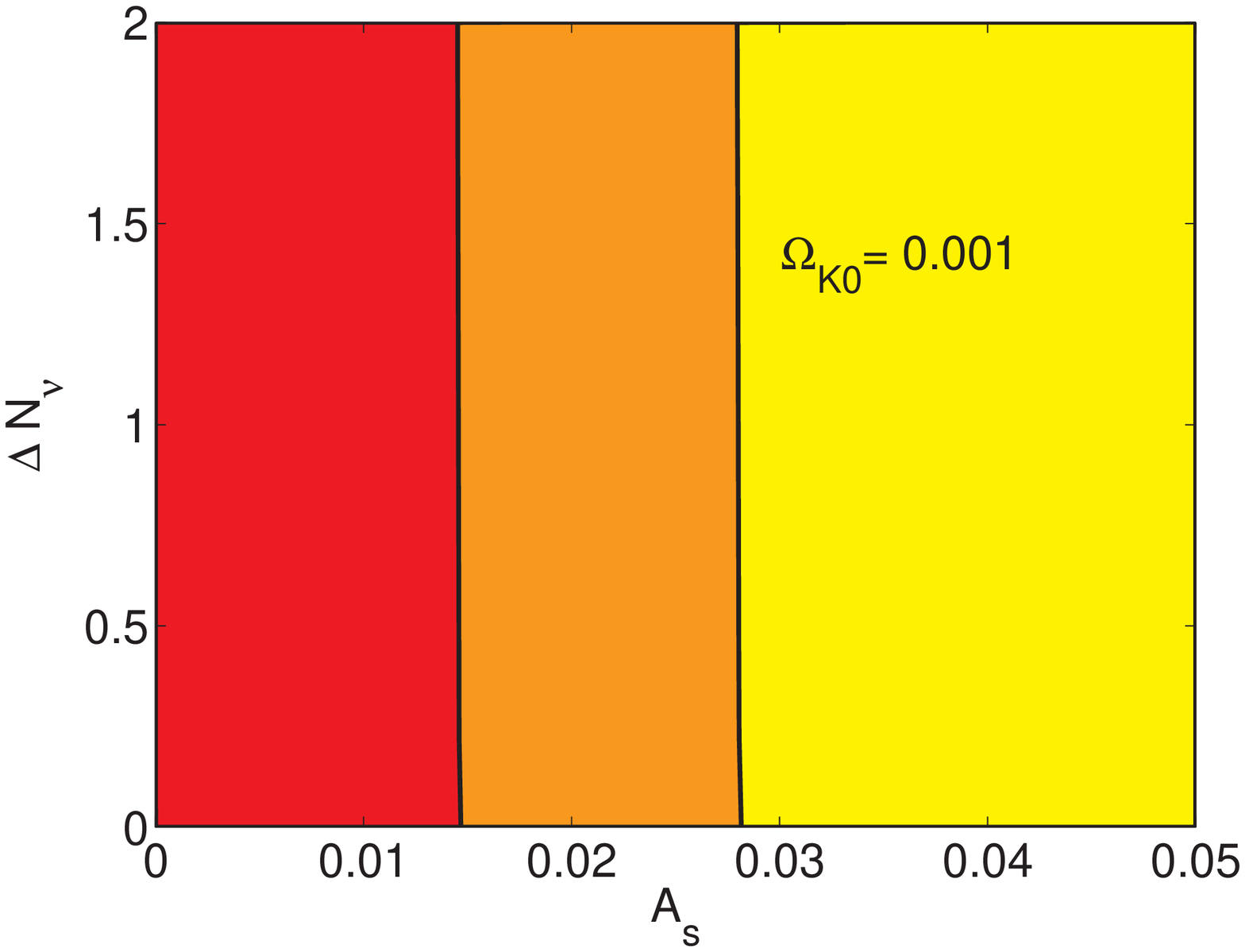,width=6
cm}&\epsfig{file=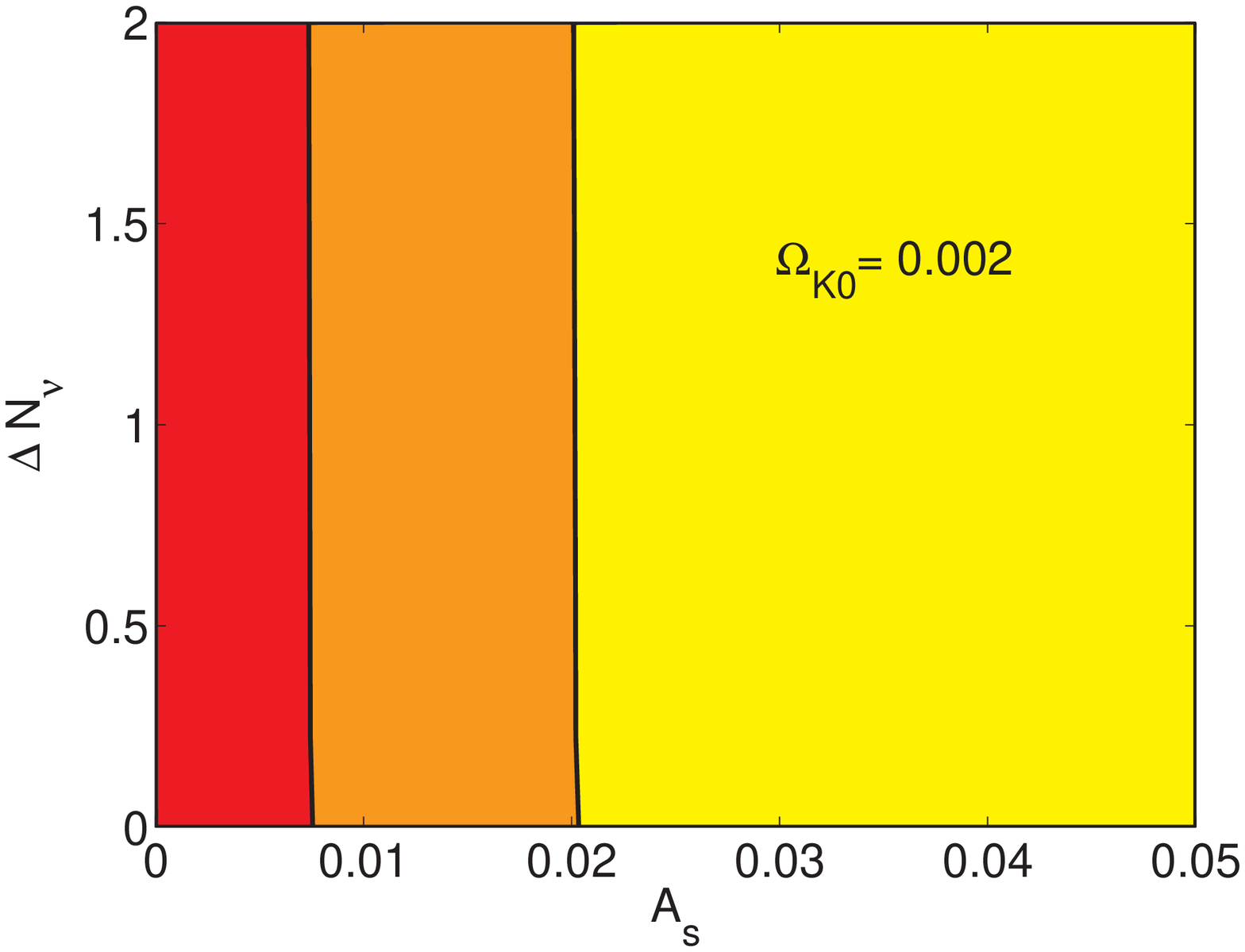,width=6
cm}
\end{tabular}
\caption{(Color online) {\it{\textbf{Beyond Detailed Balance:}
Likelihood contours of $A_s$ and $\dn$ for four choices of
$\Omega_{K0}$, with the other parameters fixed to their best-fit
values. Color scheme as in Fig. \ref{dbcontours}}.}}
\label{cont_varying_omega}
\end{figure*}
\end{center}
The present scenario involves the following parameters:
  $H_0$, $\Omega_{b0}$, $\Omega_{c0}$, $\Omega_{r0}$, $\Omega_{K0}$, $\w_1$, $\w_3$
and $\omega_4$, $A_s$ and $\beta$. Similarly to the
detailed-balance case,  we fix the parameters
$\Omega_{m0}$,$\Omega_{b0}$, $H_0$ and $\Omega_{r0}$ at their
7-year WMAP best-fit values \cite{Komatsu:2010fb}. Thus, the
remaining free parameters are $\Omega_{K0}$, $\w_1$, $\w_3$ and
$\omega_4$, $A_s$ and $\beta$, which are subject to the same two
constraints as discussed in the previous subsection. The first one
arises from the Friedman equation at $z=0$, which leads to
  \be
     \label{ndbcond1}
\Omega_{m0}+\Omega_{r0}+\Omega_{K0}+\w_1+\w_3+\w_4=1.
  \ee
 We use this constraint to eliminate the parameter $w_1$.

The second constraint arises from BBN considerations. The term
involving $\w_3$ represents the usual dark-radiation component. In
addition, the $\w_4$-term represents a kination-like component (a
quintessence field dominated by kinetic energy
\cite{kination,kination1}), also called a ``stiff fluid''. If $\dn$ represents the BBN upper
limit on the total energy density of the universe beyond standard
model constituents, then as shown in the Appendix of
\cite{Dutta:2009jn},  we have the following constraint at the time
of BBN ($z=z_{\rm BBN}$)
\cite{BBNrefs,BBNrefs1,BBNrefs2,Malaney:1993ah}:
 \be
\label{ndbcond2} \w_3+\w_4\(1+z_{\rm
BBN}\)^2=\w_{3\text{max}}\equiv0.135\dn\Omega_{r0}. \ee
  It is clear that BBN imposes an extremely strong constraint on
$\w_4$, since its largest possible value (corresponding to
$\w_3=0$) is $\sim 10^{-24}$. This is in agreement with precision constraints on stiff fluid densities at the time of BBN derived in \cite{BobSourishBBN}. Finally, $\w_{3\text{max}}$ denotes
the upper limit on $\w_3$. In the following, we use expression
(\ref{ndbcond2}) to eliminate $\w_4$. For convenience, instead of
$\w_3$ we define a new parameter
\begin{equation}
\alpha\equiv  \frac{\w_3}{\w_{3\text{max}}},
  \end{equation}
 which has the interesting physical meaning of denoting
the ratio of the energy density of the  Ho\v{r}ava-Lifshitz dark
radiation to the total energy density of Ho\v{r}ava-Lifshitz dark
radiation and kination-like components at the time of BBN.

In summary, in the scenario at hand we have the following five
free parameters: $\dn$, $\Omega_{K0}$, $\alpha$, $A_s$ and
$\beta$. We use SNIa, BAO and CMB data to perform a likelihood
analysis, and we construct the likelihood  contours for the
parameters $A_s$ and $\dn$ for various fixed choices of the other
parameters $\alpha$, $\beta$ and $\Omega_{K0}$.
\begin{figure}[ht]
\begin{center}
\epsfig{file=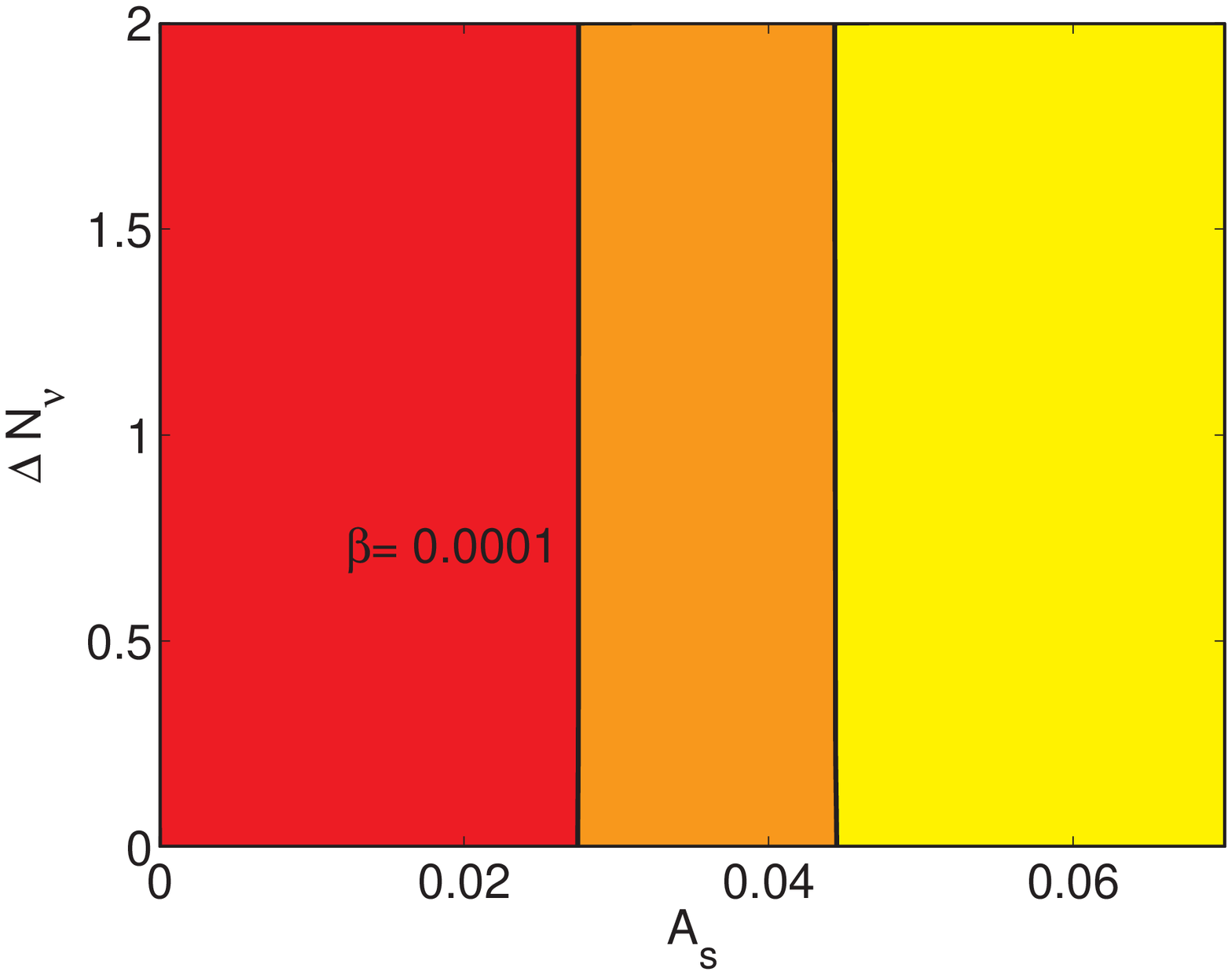,width=6
cm}\\
\epsfig{file=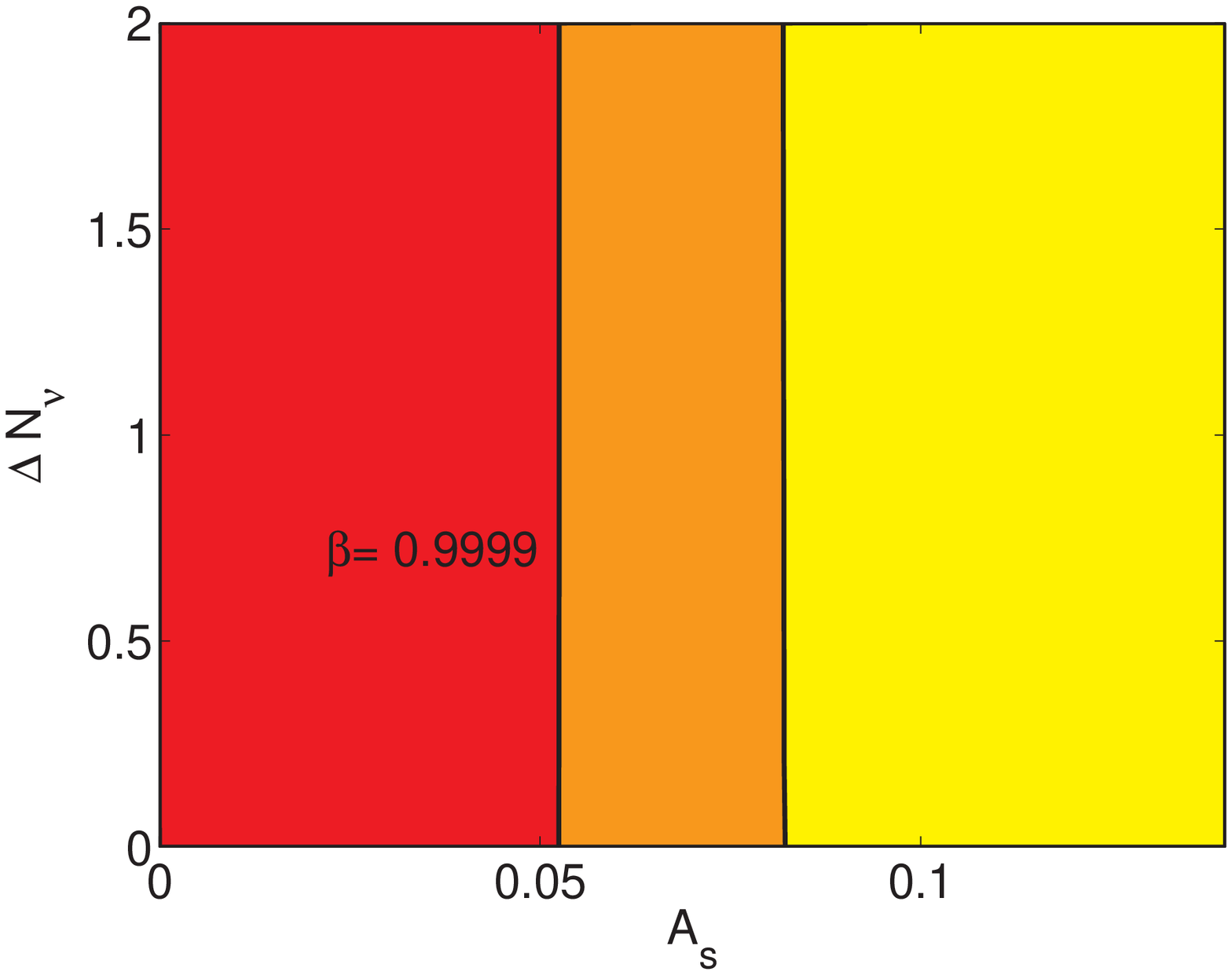,width=6 cm}
\caption{(Color online) {\it{\textbf{Beyond Detailed Balance:}
Likelihood contours of $A_s$ and $\dn$ for two extreme different
choices of $\beta$, with the other parameters fixed to their
best-fit values. Color scheme as in Fig. \ref{dbcontours}}.}}
 \label{cont_varying_beta}
\end{center}
\end{figure}

In Fig. \ref{cont_varying_alpha} we present the contours of $A_s$ and
$\dn$ obtained by varying $\alpha$ around its best-fit value,
while keeping $\beta$ and $\Omega_{K0}$ fixed at their best-fit
values. We find that as $\alpha$ increases, the
contours shrink and rotate counter-clockwise, placing tighter
constraints on $\dn$.

In Fig. \ref{cont_varying_omega} we present the contours of $A_s$
and $\dn$ for four different choices of spatial curvature, with
$\alpha$ and $\beta$ fixed at their best-fit values. We observe that curvature
constraints are quite tight. For example the 1-$\sigma$ regions shrink as
$|\Omega_{k0}|$ increases from 0.001
to 0.002 (as shown in the figure) and they almost completely disappear
for $|\Omega_{k0}|\gtrsim0.003$ (which is not shown).

In Fig. \ref{cont_varying_beta} we show the likelihood contours of
$A_s$ and $\dn$, with $\alpha$ and $\Omega_{K0}$ fixed to their
best-fit values, and two extreme choices of $\beta$.  We deduce
that the contours expand slightly for the larger value of $\beta$.

Finally, the best-fit values, along with the corresponding
$1\sigma$ confidence intervals are presented in Table
\ref{ndb_bestfits}.
\begin{table}[ht]
    \centering
        \begin{tabular}{|c|c|}
        \hline
        $A_s$           & $2.47\times10^{-10};\,\(0,2\)$                    \\\hline
        $\dn$             &  $5.23\times10^{-4};\,\(0,2.0\)$                   \\\hline
        $\alpha$         &   $5.17\times10^{-4};\,\(0,1\)$                   \\\hline
        $\Omega_{K0}$    &    $1.29\times10^{-3};\,\(-0.003,0.003\)$                  \\\hline
        $\beta$             & $4.89\times10^{-5};\,\(0,1\)$             \\\hline
        \end{tabular}
    \caption{Best-fit values and $1\sigma$ confidence intervals for the free parameters
    $A_s$, $\dn$, $\alpha$, $\Omega_{K0}$ and $\beta$, in
    the beyond-detailed-balance scenario,
    arising from a likelihood analysis using SNIa, BAO and CMB data.}
    \label{ndb_bestfits}
\end{table}

\section{Cosmological implications}
\label{cosmimpl}

In the previous sections we constructed cosmological scenarios of
a generalized Chaplygin gas in a universe governed by
Ho\v{r}ava-Lifshitz, and we used observational data in order to
impose constraints on the model parameters. 
In recent past, people have discussed different cosmological scenarios
in Ho\v{r}ava-Lifshitz gravity. Maeda et al \cite{maeda 2010} have recently
studied the bouncing as well as oscillating universe in Ho\v{r}ava-Lifshitz
gravity in presence of both curvature and cosmological constant. One very
interesting feature of their study is the possibility of a quantum tunneling from the
oscillating spacetime to an inflationary scenario. In another interesting work,
Bertolami and Zarro \cite{berto 2011} have discussed quantum cosmology via
minisuperspace model for projectable Ho\v{r}ava-Lifshitz gravity without detailed
balanced condition. They considered the presence of a cosmological constant and when it is positive they recovered the classical GR solution with accelerating expansion. In the present section
we discuss some of the cosmological implications in our case where we consider the
presence of generalised Chaplygin gas . In particular, we focus on the evolution of the
dimensionless expansion rate defined in (\ref{fidexprate}), and on
the evolution of the equation-of-state parameter of the total
cosmic fluid of the universe, defined as $w=p_{tot}/\rho_{tot}$,
with the total pressure and energy density given by
\be
p_{tot}=p_c+\frac{1}{3}\rho_{r}
+\frac{2}{\kappa^2}\left[\frac{K^2}{\Lambda a^4}-3\Lambda
\right]
\ee
\be
\rho_{tot}=\rho_c+\rho_b+\rho_{r}
+\frac{2}{\kappa^2}\left[\frac{3K^2}{\Lambda
a^4}+3\Lambda\right]
\ee
for the detailed-balanced scenario, and
\be
p_{tot}=p_c+\frac{1}{3}\rho_{r}
+\left[-\frac{\sigma_1}{6\sigma_0}+\frac{\sigma_3
K^2}{18\sigma_0a^4} +\frac{\sigma_4 K}{6\sigma_0a^6}\right]
\ee
\begin{eqnarray}
\rho_{tot}=\rho_c+\rho_b+\rho_{r}
+\left[\frac{\sigma_1}{6\sigma_0}+\frac{\sigma_3
K^2}{6\sigma_0a^4} +\frac{\sigma_4 K}{6\sigma_0a^6}\right],
\end{eqnarray}
for the beyond-detailed-balance one, as it can be easily extracted
from the corresponding Friedmann equations. Therefore, replacing
the scale factor by the redshift, and using the density parameters
(\ref{densparam}) at present, as well as the Hubble parameter as a
function of the redshift (from relations (\ref{Frdbfinal}) and
(\ref{Frndbfinal}), for the detailed and beyond detailed-balance
case respectively), we straightforwardly acquire the total
equation-of-state parameter as a function of the redshift $w(z)$.

\subsection{Detailed Balance}

\begin{center}
\begin{figure}[ht]
\epsfig{file=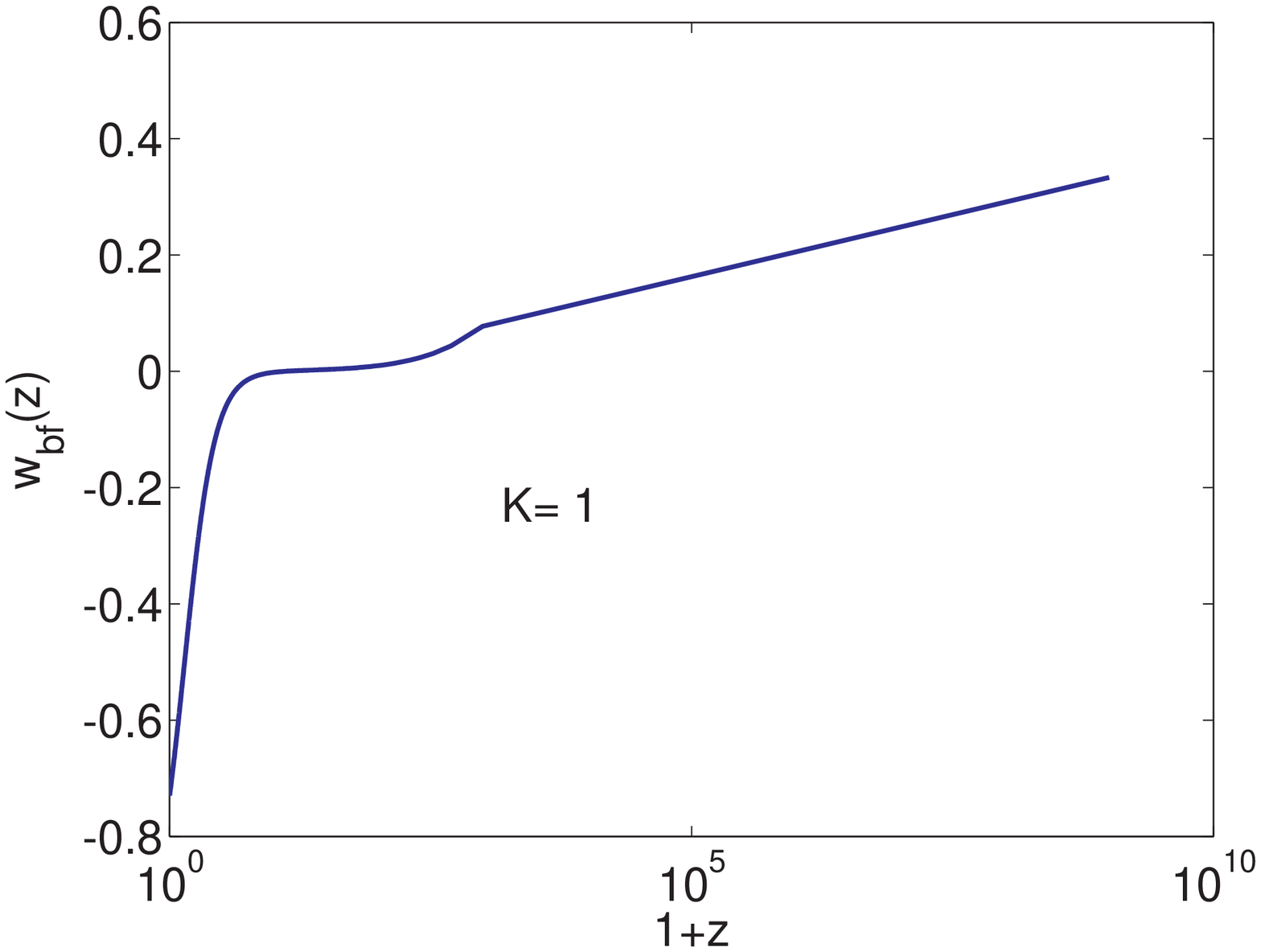,width=6
cm}\\
\epsfig{file=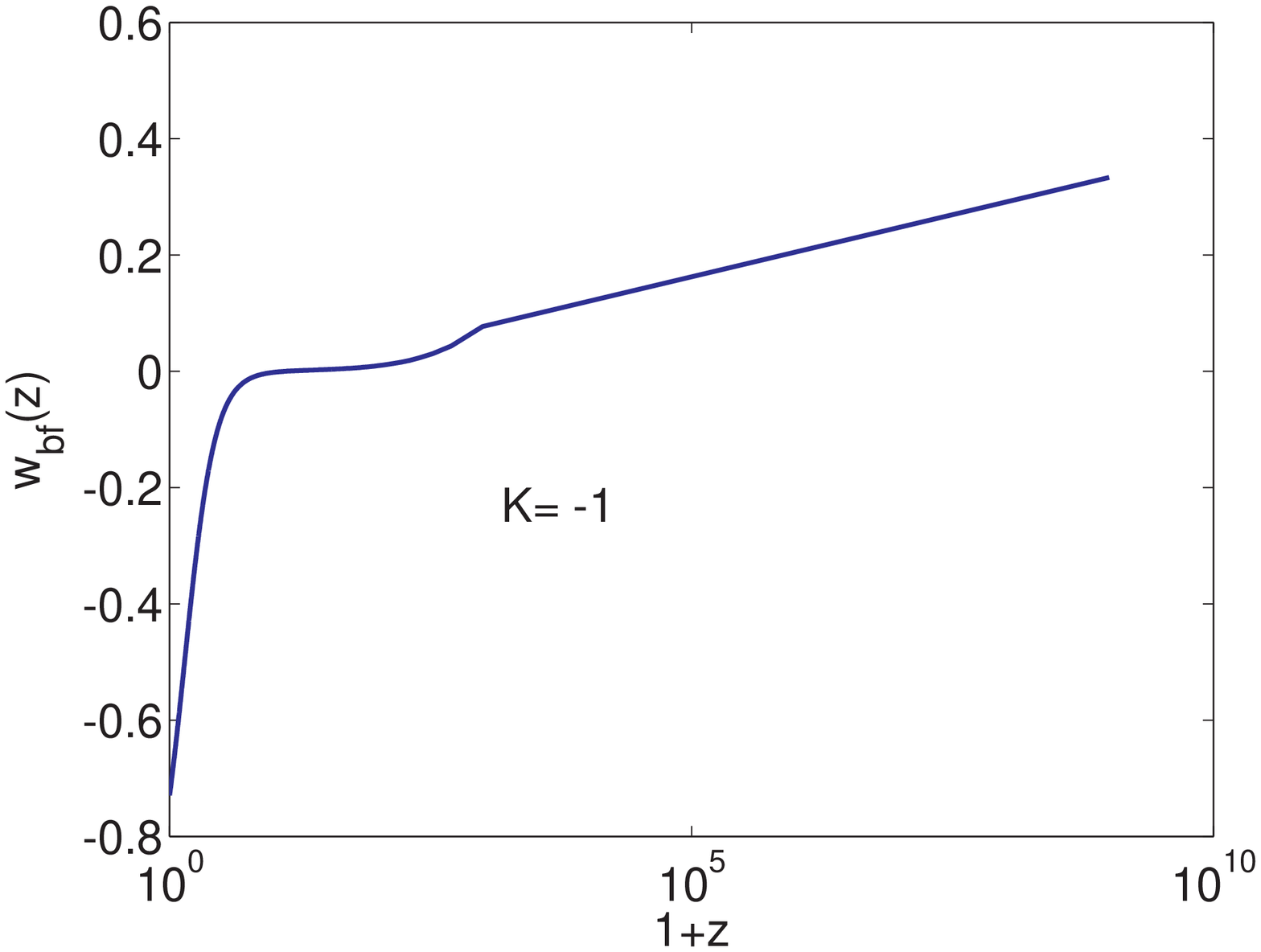,width=6 cm} \caption{(Color online)
{\it{\textbf{Detailed Balance:}  Evolution of the fiducial total
equation-of-state parameter of the cosmic fluid $w_{bf}(z)$, for
positive and negative curvatures, with all parameters fixed to
best-fit values presented in Table \ref{db_bestfits}}.}}
\label{db_w_bf}
\end{figure}
\end{center}

Let us first examine the evolution of the total equation-of-state
parameter $w(z)$ of the universe. In order to make our
presentation simpler we will use the  best-fit value set of Table
\ref{db_bestfits} as our fiducial choice for the various model
parameters, denoting by $w_{bf}(z)$ the corresponding $w(z)$. That
is, the subscript ``bf'' for a parameter will imply the best-fit
value of the parameter and for a variable will imply that all
parameters determining that variable have been fixed to their
best-fit values. In Fig. \ref{db_w_bf} we present the evolution of
$w_{bf}(z)$.

As we observe, at very early times the cosmic equation of state is
close to $1/3$, since the standard-model radiation and the dark
radiation from the Ho\v{r}ava-Lifshitz gravitational background
dominate, as expected. At intermediate redshifts, the Chaplygin
gas dominates and it behaves like a dust for quite a long time,
leading to an equation of state close to zero. Finally at late
times, the de Sitter phase of the Chaplygin gas dominates.
Interestingly, this evolution is consistent with the thermal
history of our universe, with the succession of the radiation,
matter, and ``dark energy'' epochs, and this feature acts as an
advantage of the scenario at hand. Furthermore, note that we do
not need any additional mechanism to describe the late-time
universe acceleration (that is the dark energy), since this is
obtained through a unified way by the generalized Chaplygin gas.
That is, this crucial property of the Chaplygin gas is still
exhibited by the present Ho\v{r}ava-Lifshitz embedded scenario.
\begin{center}
\begin{figure}[!]
\epsfig{file=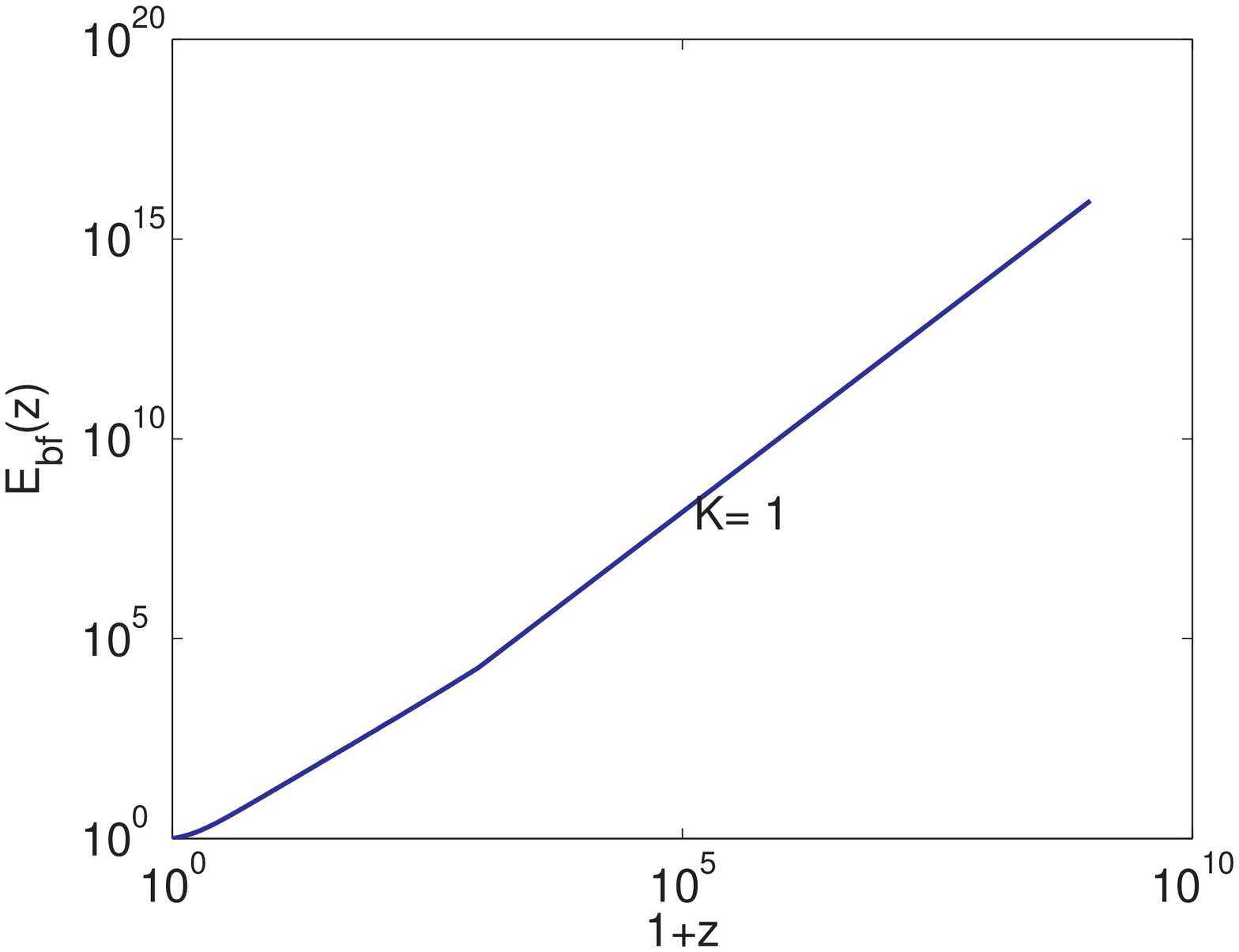,width=6
cm}\\
\epsfig{file=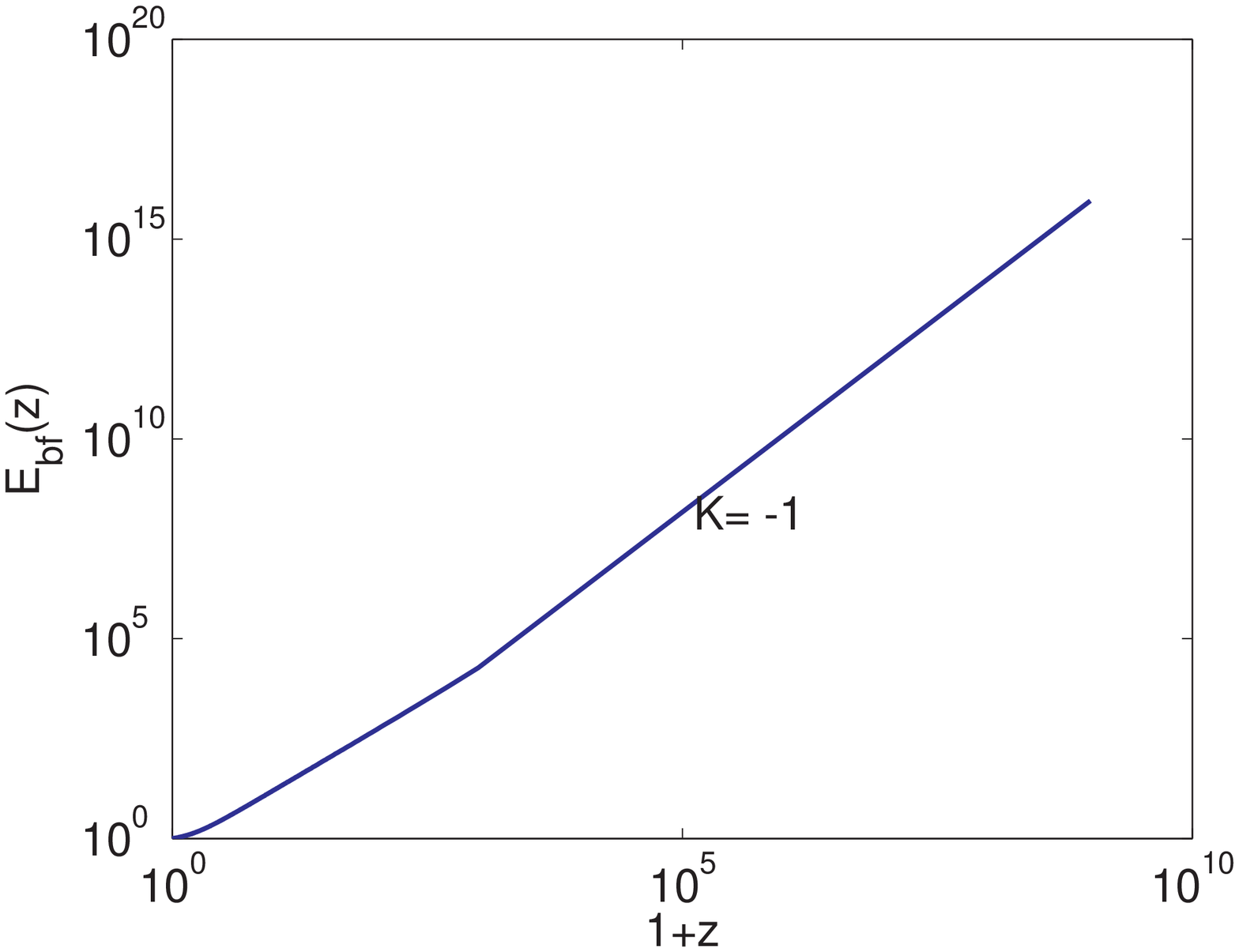,width=6 cm} \caption{(Color online)
{\it{\textbf{Detailed Balance:} Evolution of the fiducial
dimensionless Hubble parameter $E_{bf}(z)$ for positive and
negative curvatures with all parameters fixed to best-fit values
presented in Table \ref{db_bestfits}}.}}
 \label{db_E_bf}
\end{figure}
\end{center}
\begin{center}
\begin{figure*}[!]
\begin{tabular}{c@{\qquad}c}
\epsfig{file=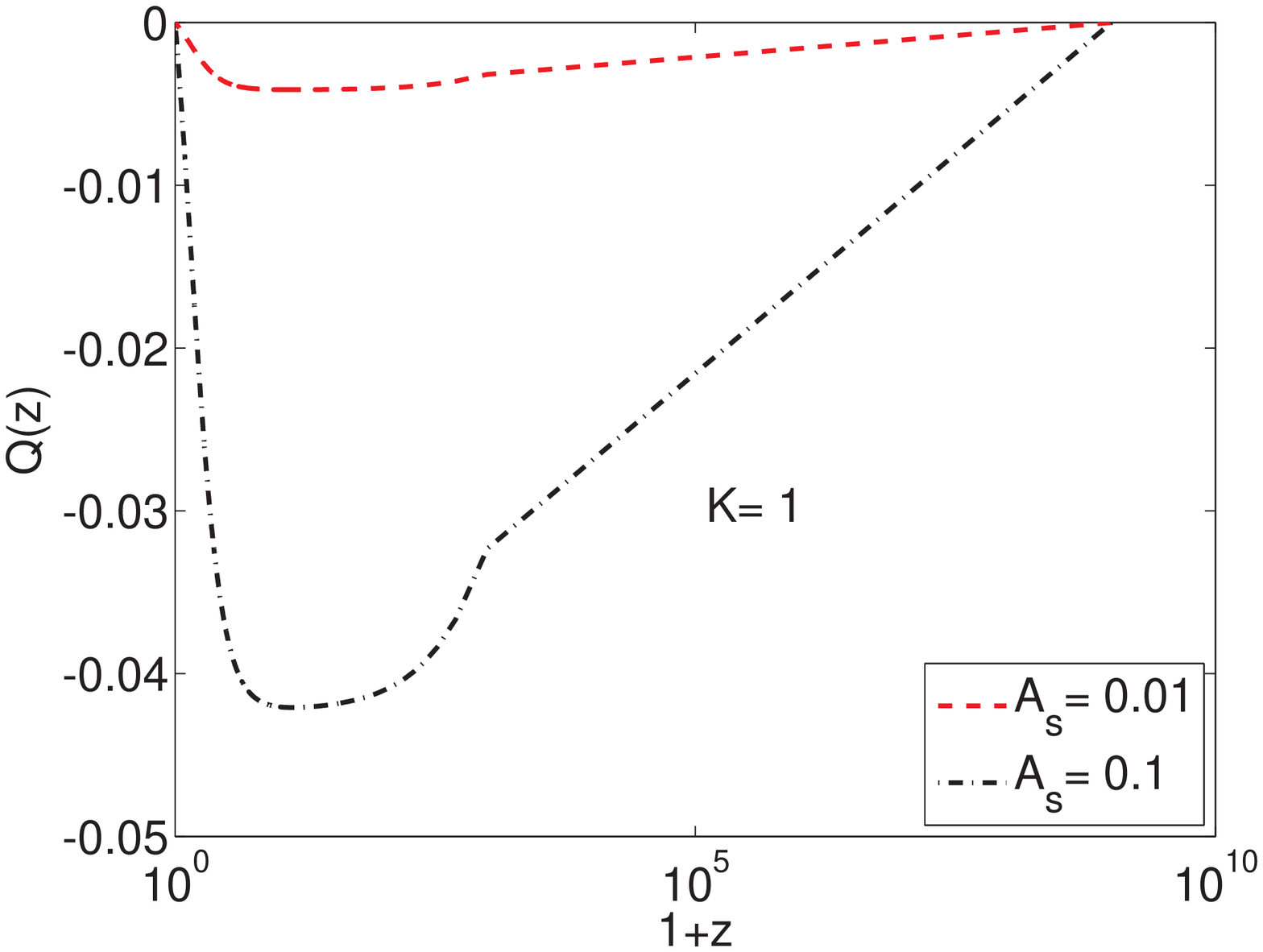,width=6 cm}&\epsfig{file=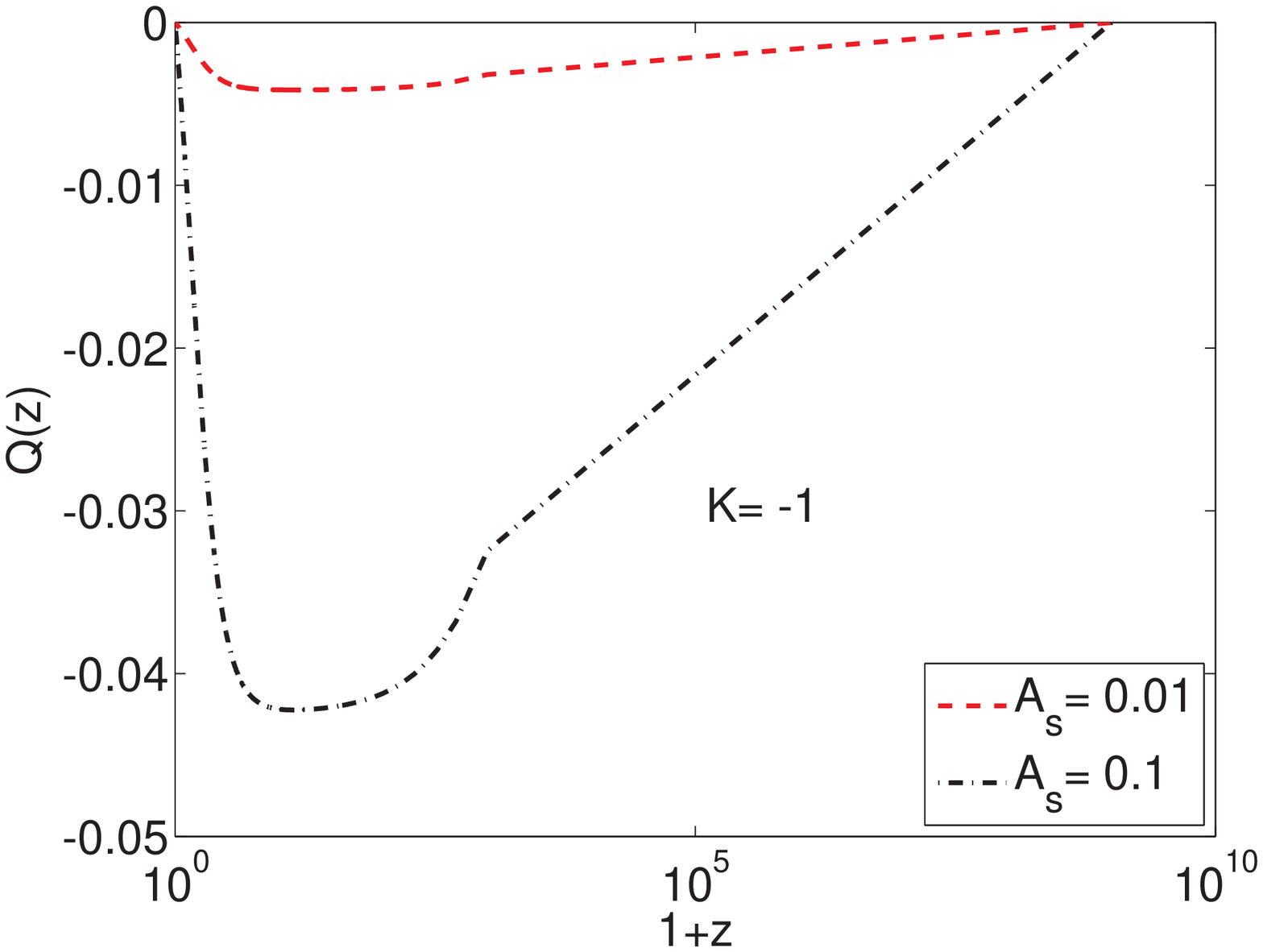,width=6 cm}\\
\epsfig{file=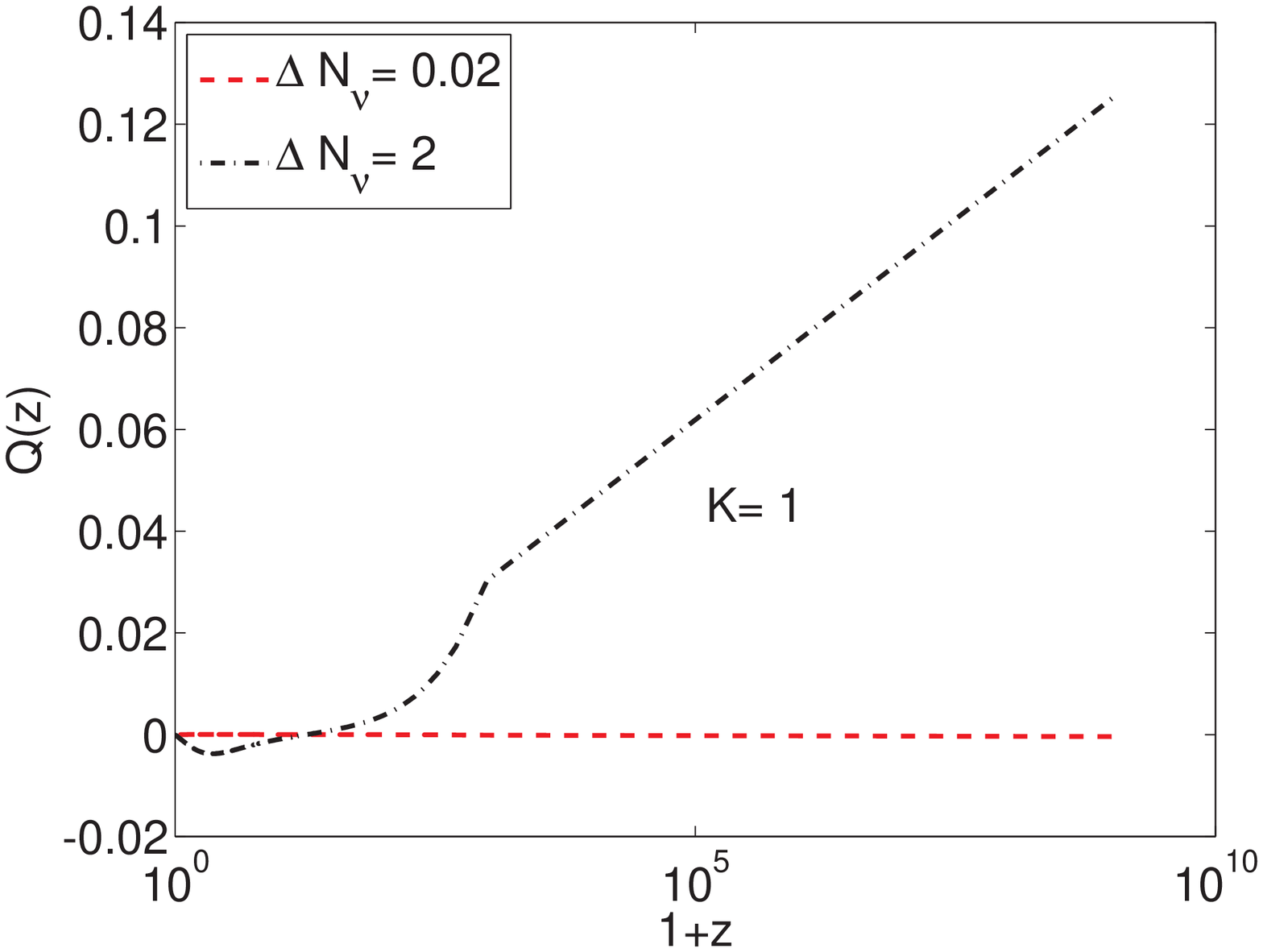,width=6 cm}&\epsfig{file=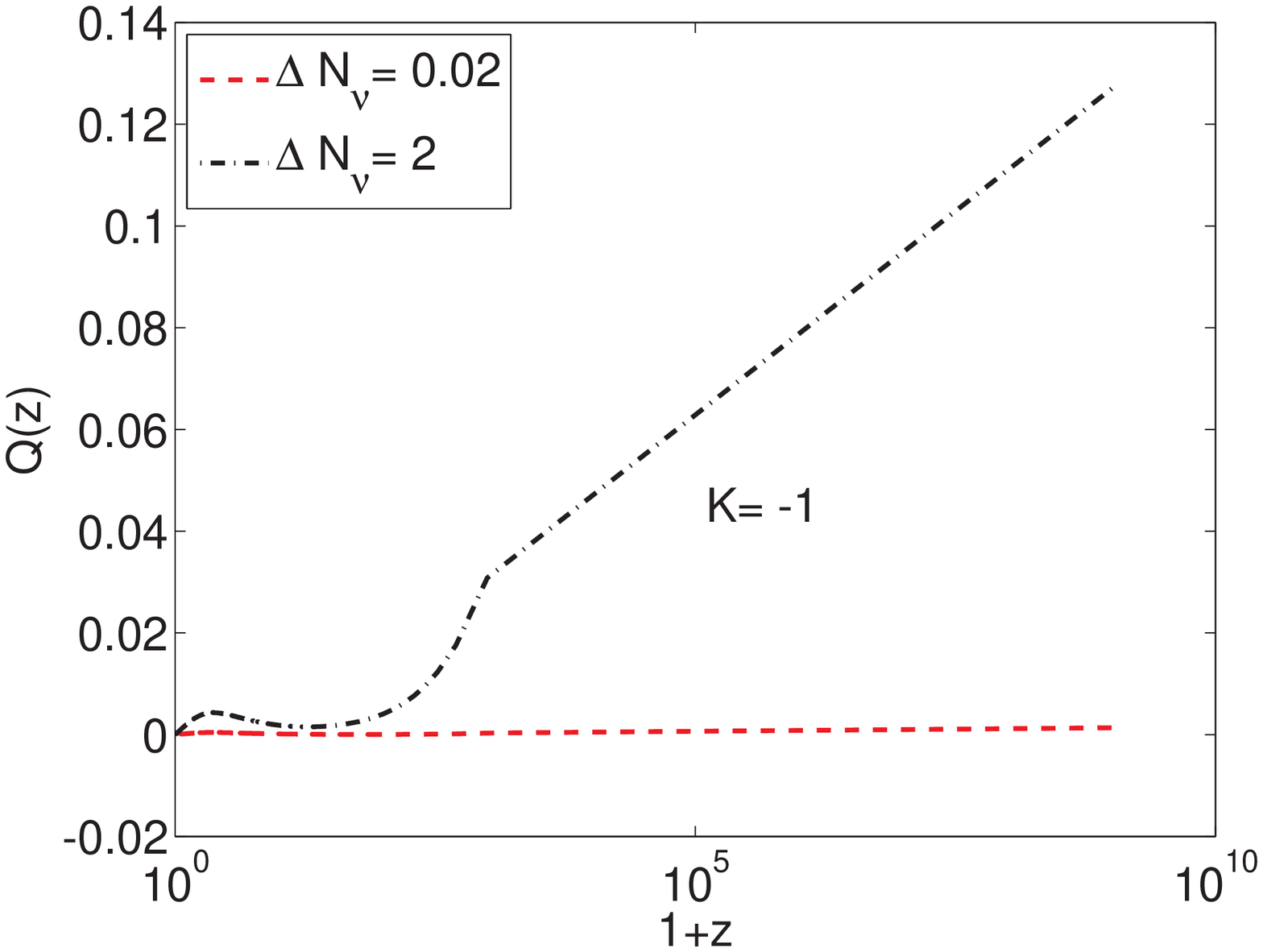,width=6 cm}\\
\epsfig{file=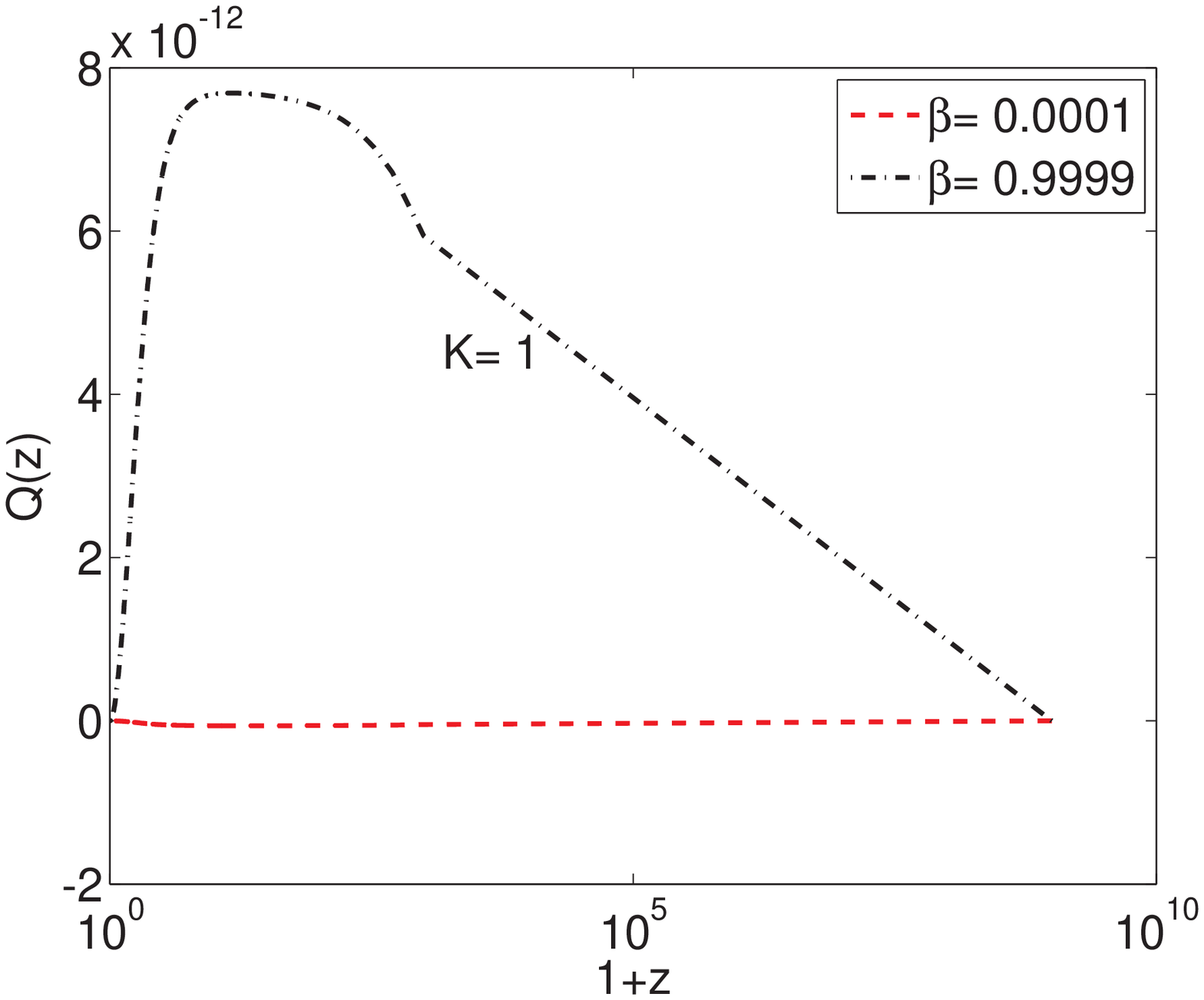,width=6
cm}&\epsfig{file=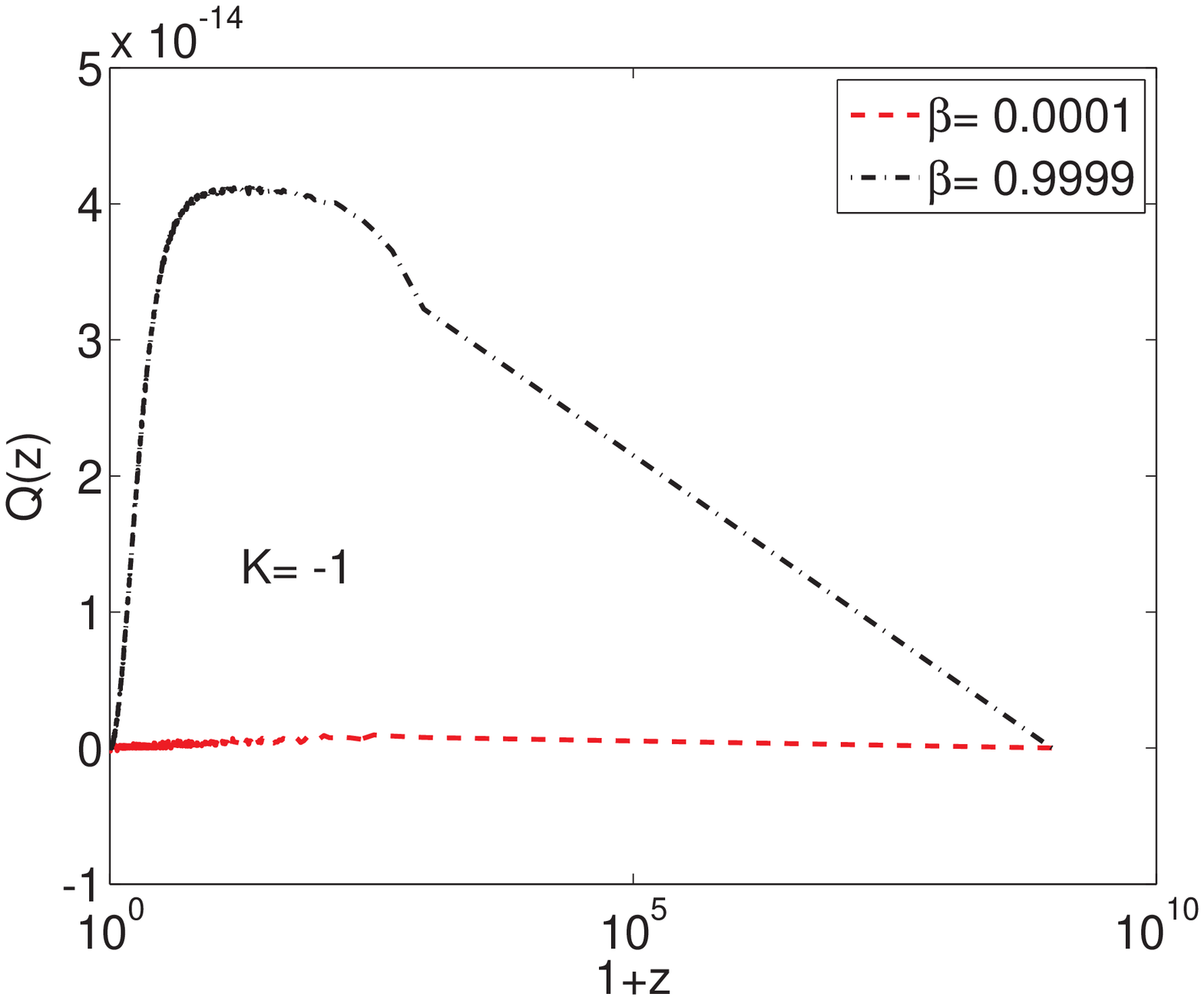,width=6 cm}
\end{tabular}
\caption{ (Color online) {\it{\textbf{Detailed Balance:}
Fractional variation $Q(z)$ of the dimensionless Hubble parameter
$E_{bf}(z)$, as each of the parameters $A_s$, $\dn$ and $\beta$
varies within its allowed limits as indicated in the graphs.}}}
 \label{db_variations}
\end{figure*}
\end{center}

We now focus on the evolution of  the dimensionless expansion rate
$E(z)$ defined in (\ref{fidexprate}). In order to acquire the
basic information, in Fig. \ref{db_E_bf} we depict the evolution
of the fiducial expansion rate $E_{bf}(z)$, that is with all
parameters fixed to their best-fit values presented in Table
\ref{db_bestfits}. As expected,  $E_{bf}(z)$ starts from very
large values, and it steadily decreases towards its present value
$1$.

It would be interesting to investigate how the expansion rate is
affected by each parameter. In order to quantify this, we define
the quantity $Q(z)$, which is the fractional change in the
expansion rate $E_{bf}(z)$ as each parameter is varied within its
allowed limits, while all the other parameters are kept fixed at
their best-fit values, that is
\begin{equation}
Q(z)\equiv\frac{E(z)-E_{bf}(z)}{E_{bf}(z)}.
\end{equation}
Note that $A_s$ and $\beta$ have well-defined upper and lower
limits, while for $\dn$ we use the standard BBN upper limit of
$\dn<2.0$ \cite{BBNrefs}. In Fig. \ref{db_variations} we present
$Q(z)$ for all the parameters of the scenario. As we see, both
$A_s$ and $\dn$ can have a strong impact on the expansion rate,
while $\beta$ has a very weak impact. The fact that the effect of
$\beta$ is small was expected, since in the contour plots of Fig.
\ref{dbcontours} we did not obtain a significant change upon
varying $\beta$.

Finally, from Fig. \ref{db_variations} we deduce that the impact
of both the generalized Chaplygin gas parameters $A_s$ and $\beta$
is realized mainly at low redshifts, while it weakens at high
ones. This is explained by the fact that at high redshifts the
Chaplygin gas is subdominant and thus the expansion dynamics is
driven by the (standard-model and dark) radiation. On the other
hand, $\dn$, which is directly related to the amount of dark
radiation, can have a significant impact on the expansion rate at
arbitrarily high redshifts.

\subsection{Beyond Detailed Balance}

Let us now repeat the analysis of the previous subsection, in the
beyond-detailed-balance version of the examined scenario.

\begin{figure}[!]
\begin{center}
\includegraphics[width=6cm]{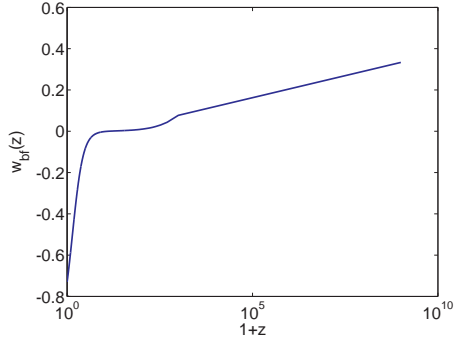}
\caption{(Color online) {\it{\textbf{Beyond Detailed Balance:}
Evolution of the fiducial total equation-of-state parameter of the
cosmic fluid $w_{bf}(z)$, with all parameters fixed to best-fit
values presented in Table \ref{ndb_bestfits}.}} }\label{w_bf}
\end{center}
\end{figure}

In Fig. \ref{w_bf} we depict the evolution of $w_{bf}(z)$, that is
the total equation-of-state parameter of the universe, with all
the model parameters fixed to their best-fit values presented in
Table \ref{ndb_bestfits}. As we observe, the behavior of
$w_{bf}(z)$ is similar to that of the detailed-balance case, that
is we obtain an initial radiation era, a relatively long matter
one, and finally an era of accelerated expansion.

\begin{figure}[!]
\begin{center}
\includegraphics[width=6cm]{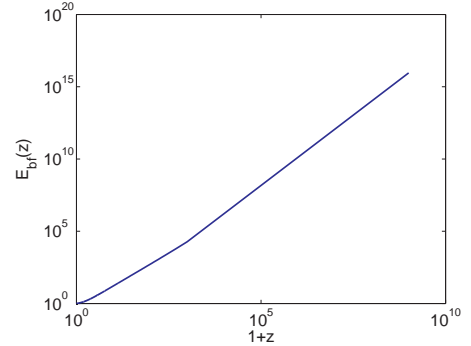}
\caption{(Color online) {\it{\textbf{Beyond Detailed Balance:}
Evolution of the fiducial dimensionless Hubble parameter
$E_{bf}(z)$, with all parameters fixed to their best-fit values
presented in Table \ref{ndb_bestfits}.}} }\label{E_bf}
\end{center}
\end{figure}

Concerning the dimensionless expansion rate of the universe
$E(z)$, in Fig. \ref{E_bf} we depict its fiducial value
$E_{bf}(z)$ as a function of the redshift. Again we see that
$E_{bf}(z)$ starts from very large values and it steadily
decreases, compatibly with observations.
\begin{figure}[!]
\begin{center}
\includegraphics[width=6cm]{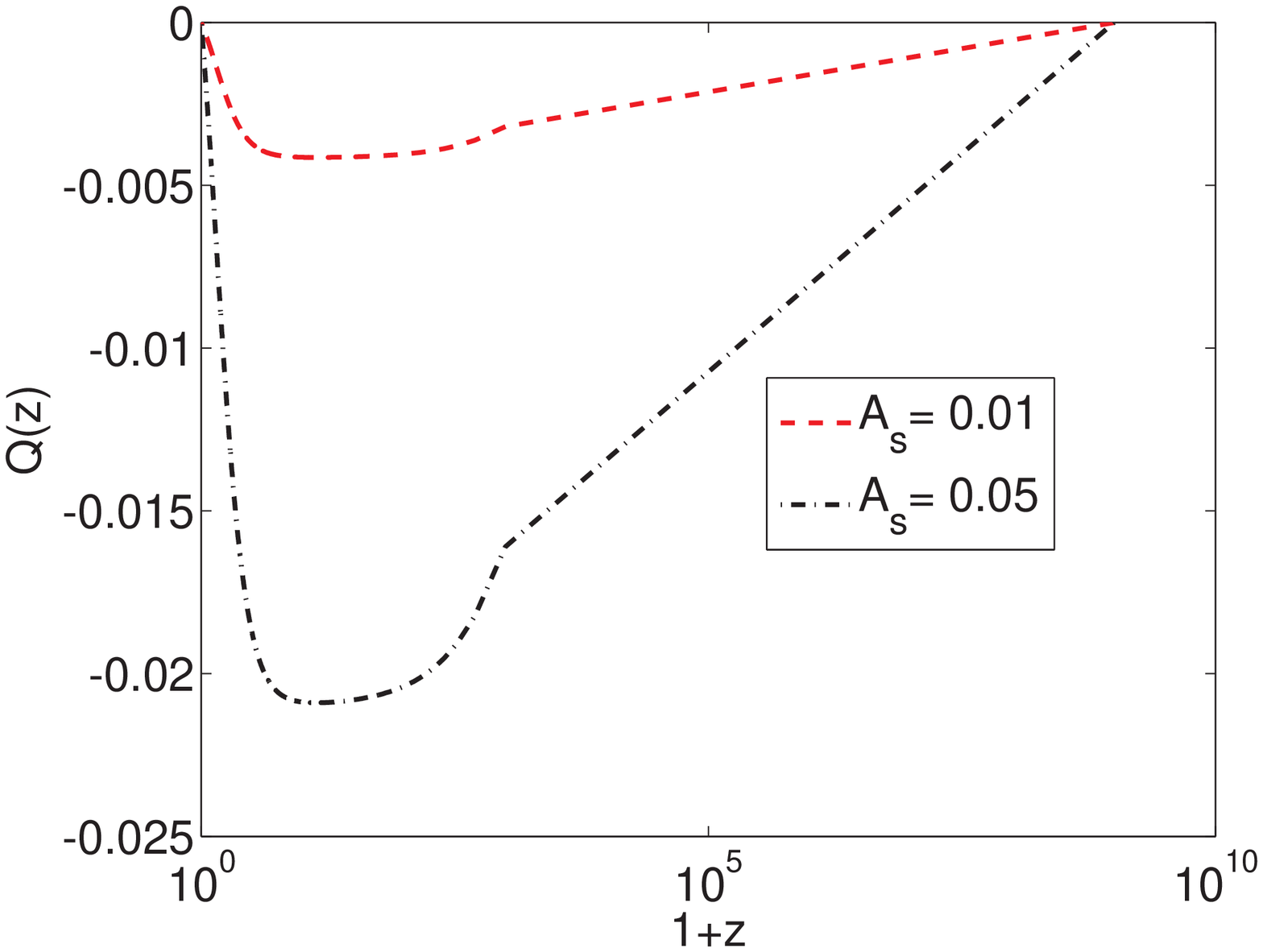}
\caption{(Color online) {\it{\textbf{Beyond Detailed Balance:}
Fractional variation $Q(z)$ of the dimensionless Hubble parameter
$E_{bf}(z)$, upon varying the parameter $A_s$ within its allowed
limits, as indicated in the graph. All the other parameters are
kept fixed to their fiducial values presented in Table
\ref{ndb_bestfits}.}}} \label{H_varying_As}
\end{center}
\end{figure}
\begin{figure}[!]
\begin{center}
\includegraphics[width=6cm]{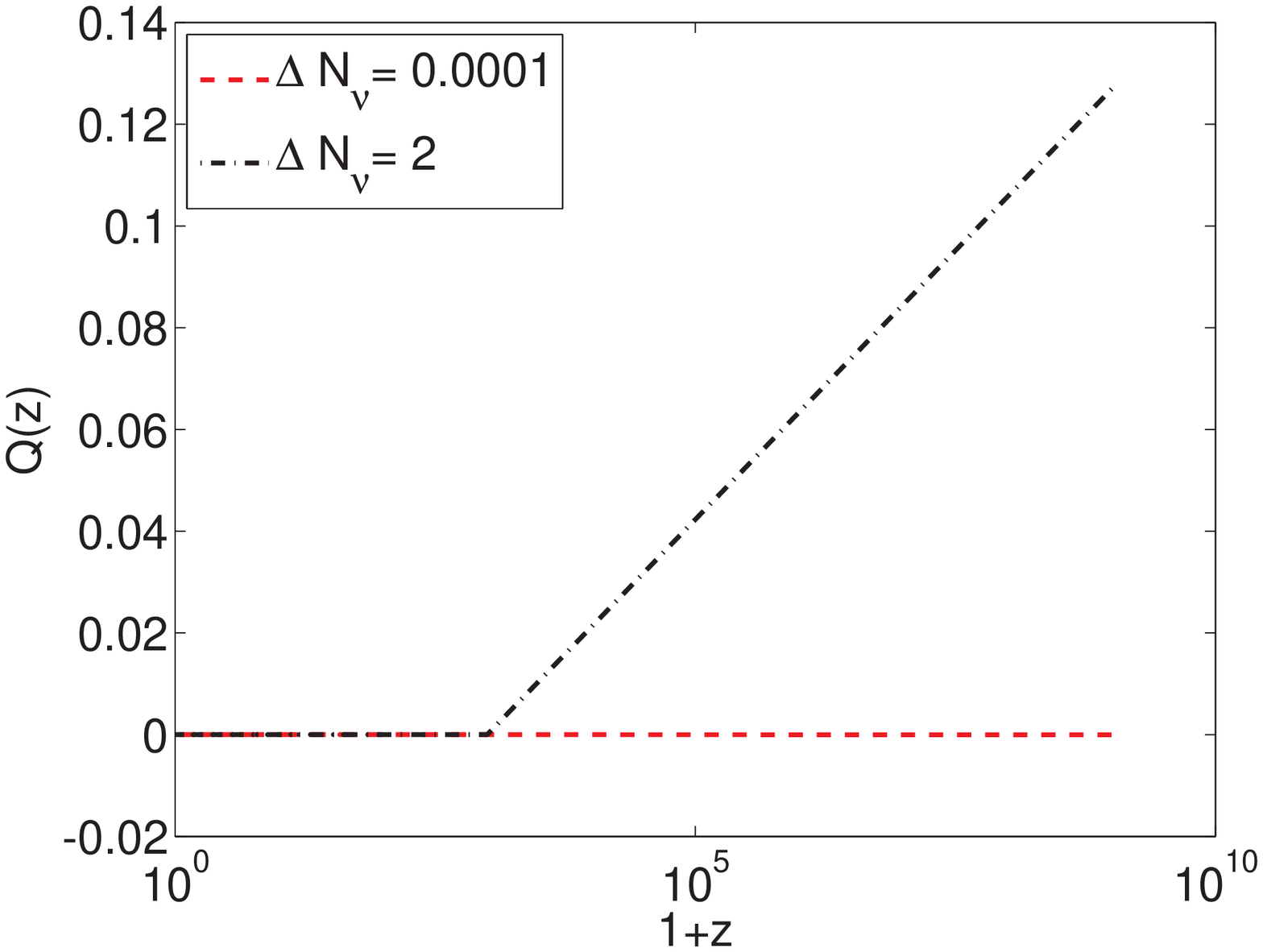}
\caption{ (Color online) {\it{ \textbf{Beyond Detailed Balance:}
Fractional variation $Q(z)$ of the dimensionless Hubble parameter
$E_{bf}(z)$, upon varying the parameter $\dn$ within its allowed
limits, as indicated in the graph. All the other parameters are
kept fixed to their fiducial values presented in Table
\ref{ndb_bestfits}.}}} \label{H_varying_Dn}
\end{center}
\end{figure}
\begin{figure}[!]
\begin{center}
\includegraphics[width=6cm]{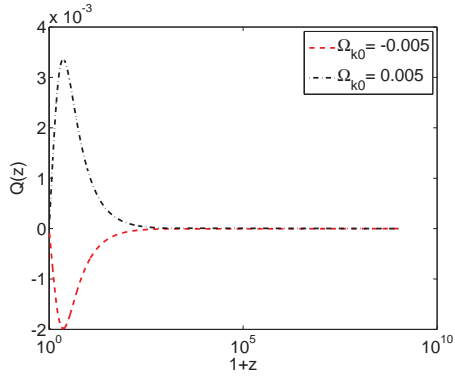}
\caption{(Color online) {\it{\textbf{Beyond Detailed Balance:}
Fractional variation $Q(z)$ of the dimensionless Hubble parameter
$E_{bf}(z)$, upon varying the parameter $\Omega_{K0}$ as indicated
in the graph. All the other parameters are kept fixed to their
fiducial values presented in Table \ref{ndb_bestfits}.}}}
\label{H_varying_Omk}
\end{center}
\end{figure}

\begin{figure}[ht]
\begin{center}
\includegraphics[width=6cm]{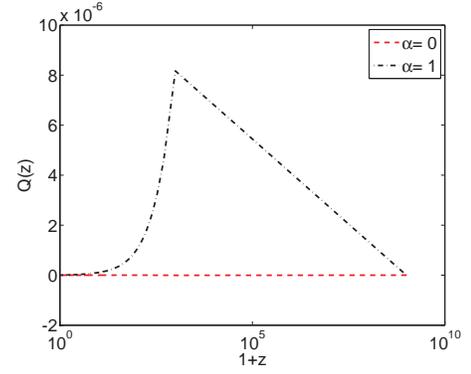}
\caption{(Color online) {\it{ \textbf{Beyond Detailed Balance:}
Fractional variation in the dimensionless Hubble parameter
$E_{bf}(z)$, upon varying the parameter $\alpha$ within its
allowed limits, as indicated in the graph. All the other
parameters are kept fixed to their fiducial values presented in
Table \ref{ndb_bestfits}.}}} \label{H_varying_alpha}
\end{center}
\end{figure}

\begin{figure}[ht]
\begin{center}
\includegraphics[width=6cm]{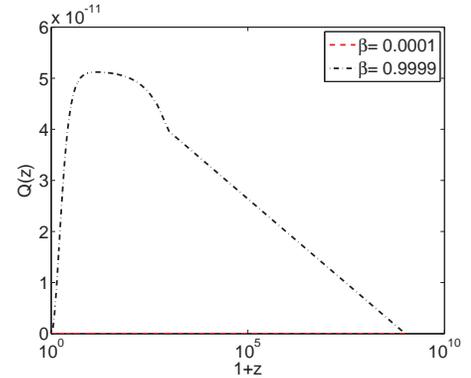}
\caption{(Color online) {\it{\textbf{Beyond Detailed Balance:}
Fractional variation in the dimensionless Hubble parameter
$E_{bf}(z)$, upon varying the parameter $\beta$ within its allowed
limits indicated, as in the graph. All the other parameters are
kept fixed to their fiducial values presented in Table
\ref{ndb_bestfits}.}}}\label{H_varying_beta}
\end{center}
\end{figure}

We now proceed to the investigation of the effect that the various
model parameters have on the expansion rate. In Figures
\ref{H_varying_As}-\ref{H_varying_beta} we demonstrate the
fractional change $Q(z)$ of the expansion rate $E_{bf}(z)$ as each
parameter is varied within its allowed limits, as it is indicated
in the graphs, while all other parameters are kept fixed at their
best-fit values. From Fig. \ref{H_varying_As} we deduce that $A_s$
can have a strong impact on the expansion rate at low redshifts,
while from Fig. \ref{H_varying_Dn}  we see that $\dn$ can affect
significantly the expansion rate at large redshifts. On the
other hand, form Figures \ref{H_varying_Omk}, \ref{H_varying_alpha} and
\ref{H_varying_beta} we observe that $\Omega_{k0}$
and especially the parameters $\alpha$ and
$\beta$ have negligible impacts. Similarly to the detailed balance
case, the fact that the effect of $\beta$ is small was expected,
since in the contour plots of Fig. \ref{cont_varying_beta} we did
not obtain a significant change upon varying $\beta$.


The explanation of the effect of the various parameters on the
expansion rate is easily obtained, knowing the behavior of the
various terms as a function of the redshift. In particular, at
high redshifts the radiation (standard-model and dark one)
dominates and drives the expansion, while the Chaplygin gas is
subdominant. However $\dn$, which quantifies  the
Ho\v{r}ava-Lifshitz dark radiation and kination-like components,
can have a significant impact on the expansion rate at arbitrarily
high redshifts.

\section{Conclusions}
\label{conclusions}

In this work we investigated the cosmological scenario of
generalized Chaplygin Gas in a universe governed by
Ho\v{r}ava-Lifshitz gravity, both in the detailed-balance as well as
in the beyond-detailed balance version of the theory. Furthermore,
in order to obtain a realistic cosmology we have included the
baryonic matter and standard-model radiation sectors.

After extracting the cosmological equations, we used data from
SNIa, BAO and CMB observations, as well as arguments from Big Bang
Nucleosynthesis, in order to impose constraints on the various
parameters of the scenario. For the  detailed-balance case we have
three free parameters, namely the exponent $\beta$ of the
generalized Chaplygin Gas, its present equation-of-state parameter
value $A_{s}$, and the effective neutrino parameter $\Delta
N_{\nu}$ which quantifies the total amount of Ho\v{r}ava-Lifshitz
dark-radiation and kination-like components allowed during BBN.
The best fit values of the parameters together with the
corresponding $1\sigma$ confidence intervals, arisen from the
likelihood analysis for open and closed universe, were shown in
Table \ref{db_bestfits}, while in Fig. \ref{dbcontours} we
presented the corresponding likelihood contours. We deduced that
the scenario at hand is compatible with observations, however the
data lead to strong bounds on $A_{s}$ and $\Delta N_{\nu}$ and the spatial curvature $\Omega_{K0}$.
 Finally, the scenario at hand depends only slightly on
$\beta$.

In the beyond-detailed-balance scenario the free parameters are
five, namely  $\beta$, $A_{s}$, $\Delta N_{\nu}$, the curvature
density $\Omega_{K0}$, and the parameter $\alpha$, which is the
ratio of the dark-radiation energy density to the sum of
dark-radiation and kination-like energy densities at the time of
BBN. The best fit values of the parameters together with the
corresponding $1\sigma$ confidence intervals, arisen from the
likelihood analysis, were shown in Table \ref{ndb_bestfits}.
Additionally, in Figures
\ref{cont_varying_alpha}-\ref{cont_varying_beta} we presented the
likelihood contours of the model parameters. From this analysis we
deduced that the entire (consistently with BBN) range of
$0\leq\Delta N_{\nu}\leq2.0$ is allowed, for suitable choices of
the other parameters. We also found tight constraints on the curvature and
on $A_s$.

After the observational elaboration, we investigated the
cosmological implications of the examined scenarios. In
particular, we focused on the evolution of the total
equation-of-state parameter of the universe, and of the expansion
rate. In the detailed-balance case, we saw that in general at very
early-times is radiation that dominates, at intermediate redshifts
is the Chaplygin gas that dominates, behaving like matter for a
long time, and finally at late times the dominant Chaplygin gas
behaves like dark energy, triggering the accelerating expansion
(see Fig \ref{db_w_bf}). This evolution is consistent with the
thermal history of the universe, and this is an advantage of the
present scenario. Note moreover that the present accelerated era
is described in a unified way, without the need of any additional
mechanism. Concerning the qualitative dependence on the various
parameters, we found that the expansion rate has a strong
dependence on both $A_{s}$ and $\Delta N_{\nu}$ at low redshifts,
while at high redshifts, where the Chaplygin gas is not dominant,
this dependence weakens and disappears (see Fig
\ref{db_variations}). Lastly, the dependence on $\beta$ is very
weak at all redshifts.

In the beyond-detailed-balance scenario we found a similar
behavior for the total equation-of-state parameter of the
universe, that is we obtained successively a radiation, a matter,
and a dark energy era (see Fig \ref{w_bf}). Qualitatively, $A_{s}$
has a weak impact on the expansion rate at
lower redshifts, $\Delta N_{\nu}$ has a significant effect at
large redshifts, while $\Omega_{K0}$ and especially $\alpha$ and $\beta$
have a negligible impact on the expansion rate at all redshifts (see
Figures \ref{H_varying_As}-\ref{H_varying_beta}).

In summary, the generalized Chaplygin gas in Ho\v{r}ava-Lifshitz
gravitational background is compatible with observations, and can
successfully reproduce the expansion history of the universe.
However, we should mention that the present analysis does not
enlighten the discussion about the possible conceptual problems
and instabilities of Ho\v{r}ava-Lifshitz gravity,  nor it can
address the questions concerning the validity of its theoretical
background, which is the subject of interest of other studies. It
just analyzes the phenomenological consequences and the
cosmological implications of the generalized Chaplygin Gas in such
a gravitational background, and thus its results can be taken into
account only if Ho\v{r}ava-Lifshitz gravity passes successfully
the necessary theoretical tests.

In the same lines, the present work does not address the problem
of undesirable instabilities and oscillations in the matter power
spectra, that GCG scenario faces in standard Einstein gravity.
Although, as discussed in the Introduction, there are several
approaches which try to solve this problem in conventional
gravity, it may be quite interesting to see whether it can be
addressed in the context of Ho\v{r}ava-Lifshitz gravity. Let us
make some comments in this perspective. As discussed in
\cite{Bento:2002ps, Bertolami:2004ic}, GCG can be modelled in
terms of scalar fields having both canonical as well as
non-canonical kinetic energy terms, and thus perturbing the GCG
one has to essentially perturb these scalar fields. In the case of
Ho\v{r}ava-Lifshitz gravity one expects in general the linear
perturbations to be quite different due to higher-order
curvature-terms present in the action. These higher-order gradient
terms can in principle generate non-adiabatic pressure
perturbation, which can help to cure the instabilities or
oscillations in the power spectra even in the large scales. The
effect can be prominent with scalar fields having non-canonical
kinetic energy terms, for instance of Dirac-Born-Infeld form,
which is typical for GCG-like equation of state.

In conclusion, the generalized Chaplygin gas scenario in
Ho\v{r}ava-Lifshitz cosmology, can have rich cosmological
consequences. The present study is a first step in this direction,
where we have considered the background evolution and have
confronted the model with a variety of currently available
observational data. The next step is to consider the inhomogeneous
GCG model in Ho\v{r}ava-Lifshitz gravity, which will be our goal
in the near future.

\begin{acknowledgments}
A.A.S acknowledges the financial support provided by the
University Grants Commission, Govt. Of India, through major
research project grant (Grant No:33-28/2007(SR)). S.D wishes to thank Ujjaini Alam for useful discussions.
\end{acknowledgments}

\end{document}